\documentclass[10pt,Journal]{IEEEtran}\IEEEoverridecommandlockouts
\usepackage{algorithmic}
\usepackage[linesnumbered,ruled]{algorithm2e}
\usepackage{amsfonts}
\usepackage{amsmath}
\usepackage{amssymb}
\usepackage{amsthm}
\usepackage{bm}
\usepackage{bbm}
\usepackage{booktabs}
\usepackage{caption}
\usepackage{CJK}
\usepackage{color}
\usepackage{cuted}
\usepackage{geometry}
\usepackage{graphicx}
\usepackage{indentfirst}
\usepackage[marginal]{footmisc}
\usepackage{multirow}
\usepackage{rotating}

\allowdisplaybreaks[4]

{\bgroup
 \addtolength\abovedisplayshortskip{#1}
 \addtolength\abovedisplayskip{#1}
 \addtolength\belowdisplayshortskip{#1}
 \addtolength\belowdisplayskip{#1}}
{\egroup\ignorespacesafterend}

\newcounter{MYtempeqncnt}

\theoremstyle{plain}

\begin{document}

\title{Joint Beamforming and Position Optimization for IRS-Aided SWIPT with Movable Antennas}

\author{Yanze Zhu, Qingqing Wu, Xinrong Guan, Ziyuan Zheng, Honghao Wang, Wen Chen, Yang Liu, and Yuan Guo \thanks{\hspace{0.35cm}Y. Zhu, Q. Wu, Z. Zheng, H. Wang, W. Chen, and Y. Guo are with the Department of Electronic Engineering, Shanghai Jiao Tong University, 200240, China, email: \{yanzezhu, qingqingwu, zhengziyuan2024, hhwang, wenchen, yuanguo26\}@sjtu.edu.cn.}\thanks{\hspace{0.35cm}X. Guan is with the College of Communications Engineering, Army Engineering University of PLA, Nanjing, 210007, China, email: guanxr@aliyun.com.}\thanks{\hspace{0.35cm}Y. Liu is with the School of Information and Communication Engineering, Dalian University of Technology, Dalian, China, and also with the National Mobile Communications Research Laboratory, Southeast University, Nanjing, China, email: yangliu\_613@dlut.edu.cn.}
\vspace{-0.2cm}}

\maketitle

\begin{abstract}
Simultaneous wireless information and power transfer (SWIPT) has been envisioned as a promising technology to support ubiquitous connectivity and reliable sustainability in Internet-of-Things (IoT) networks, which, however, generally suffers from severe attenuation caused by long distance propagation, leading to inefficient wireless power transfer (WPT) for energy harvesting receivers (EHRs). This paper proposes to introduce emerging intelligent reflecting surface (IRS) and movable antenna (MA) technologies into SWIPT systems aiming at enhancing information transmission for information decoding receivers (IDRs) and improving receive power of EHRs. We consider to maximize the weighted sum-rate of IDRs via jointly optimizing the active and passive beamforming at the base station (BS) and IRS, respectively, as well as the positions of MAs, while guaranteeing the requirements of all EHRs. To tackle this challenging task due to the non-convexity of associated optimization, we develop an efficient algorithm combining weighted minimal mean square error (WMMSE), block coordinate descent (BCD), majorization-minimization (MM), and penalty duality decomposition (PDD) frameworks. Besides, we present a feasibility characterization method to examine the achievability of EHRs’ requirements. Simulation results demonstrate the significant benefits of our proposed solutions. Particularly, the optimized IRS configuration may exhibit higher performance gain than MA counterpart under our considered scenario.
\end{abstract}


\section{Introduction}

With the rapid development of wireless communication systems, the number of Internet-of-Things (IoT) devices have experienced explosive growth in recent years to achieve pervasive intelligence and enable emerging applications such as smart cities, virtual reality and so on. To support such a massive number of wireless devices which require ubiquitous communication connectivity as well as sustainable power supply, the simultaneous wireless information and power transfer (SWIPT) technique, which employs the dual use of radio frequency (RF) signals, have attracted great attention from both academia and industry [1]. However, an energy harvesting receiver (EHR) typically requires the receive power several orders of multitude higher than that of the signal received by an information decoding receiver (IDR), since their sensitivities and applications are significantly different in practice. Hence, due to the severe attenuation caused by long distance propagation, the practical SWIPT systems are indeed bottlenecked by inefficient wireless power transfer (WPT) for EHRs. Although multiple-input-multiple-output (MIMO) technology [2] may be a promising technology to tackle this issue, the power expenditure and hardware cost may be prohibitively high for practical implementation [3].

To overcome the aforementioned drawbacks, intelligent reflecting surface (IRS), which has been envisioned as a key technology for sixth-generation (6G) wireless communications and beyond, can function as an effective solution [4]. The IRS consists of a large number of passive reflecting elements, each of which can flexibly adjust its phase shift. Via delicately configuring the phase shifts of IRS elements, the IRS can reconstruct wireless propagation environment and therefore improve the quality of emitted signals. Since none of the active components, e.g., RF-chains, are not adopted, the IRS is distinguished by negligible power consumption and low hardware complexity. In the existing literature, lots of works have demonstrated the potentials of IRS [5]-[12]. For instance, the authors of [5] proposed to deploy an IRS in a conventional communication system for reducing the transmit power at the base station (BS). Particularly, as unveiled by [5], the IRS can achieve a power scaling law of $N^{2}$ ($N$ represents the number of reflecting elements). The work [6] considered an IRS-assisted multi-cell multi-user scenario and uncovered that the IRS can significantly enhance the data transmission for the users at the cell edge. In [7], an energy efficiency (EE) maximization problem via jointly optimizing BS transmit power and IRS phase shifts was taken into account. It has been verified by [7] that the system EE can be remarkably boosted by utilizing an IRS. Thanks to its unique advantages mentioned above, the IRS has been introduced to SWIPT systems to improve their capacities [8]-[12]. Specifically, the authors of [8] considered an SWIPT system aided by an IRS, which can dramatically increase the energy harvested by EHRs. The work [9] proposed to employ multiple IRSs in an SWIPT system for reducing the transmit power at the BS. In [10], an SWIPT system including multiple IRSs with interference channels was taken into account. As reported by [10], the sum-rate performance can be drastically improved by deploying IRSs. The authors of [11] investigated a max-min EE problem in an SWIPT system with IRS implementation and demonstrated that the minimal EE of users can be significantly boosted by using an IRS. In [12], the reinforcement learning based method was exploited to enhance the resource allocation in an SWIPT system with the assistance of a sustainable IRS.

Besides IRS, movable antennas (MAs) [13], also known as fluid antennas (FAs) [14], which have been proposed very recently, can be another promising solution to the bottlenecks of SWIPT systems. Unlike traditional fixed position antennas (FPAs) without any movability, MAs can move flexibly within a specific region by adopting stepper motors [13] or liquid metal techniques [14]. Via appropriately adjusting the positions of MAs, the MA array can reconfigure wireless channel conditions, promoting signal transmission and reception. There have been various works studying the potentials of MAs in the existing literature, e.g., [15]-[21]. For instance, the authors of [15] proposed a field-response based channel model in a single-input-single-output (SISO) system, where the transmitter and receiver are equipped with one MA, respectively. As unveiled by [15], according to performance analyses, the signal-to-noise ratio (SNR) can be significantly improved by deploying MAs. The work [16] considered a MIMO system where the transmitter and receiver are empowered by multiple MAs, respectively, and uncovered that MAs can remarkably enhance the system capacity. In [17], MAs were exploited to wireless sensing systems for reducing estimation error. To further leverage the strengths of MAs, several works have introduced MAs to WPT/SWIPT systems, e.g., [18]-[21]. Specifically, the authors of [18] proposed to employ MAs in wireless-powered mobile-edge computing (MEC) systems to boost sum-computational-rate. The work [19] considered an MA-enabled wireless-powered non-orthogonal multiple access (NOMA) system, where both continuous and discrete positions of MAs were included. As reported by [19], the system throughput can be dramatically increased via implementing MAs. In [20], MAs were adopted in an SWIPT system to improve the received power at EHRs. The authors of [21] took MA-assisted SWIPT systems into account and demonstrated that the physical layer security of SWIPT systems can be drastically enhanced with MA array deployment.

Thanks to the merits brought by IRSs and MAs, some researchers have concentrated on the integration of both in communication systems to obtain higher performance gain, e.g., [22]-[28]. For instance, the authors of [22] presented some challenges and corresponding solutions of the systems with the coexistence of IRSs and MAs. The work [23] verified that the rate performance can be significantly improved by deploying both an IRS and MAs. Particularly, as unveiled by [23], the performance improvement resulting from MAs over FPAs might reduce with IRS passive beamforming being optimized. In [24], both an IRS and an MA array were employed in a multi-user system to boost the sum-rate of users. The authors of [25] proposed to utilize MAs and multiple IRSs for enhancing the coverage of considered system. The work [26] designed a secure transmission scheme for an IRS-aided integrated sensing and communication (ISAC) system, where the users are individually equipped with an MA to improve the system capacity. In [27], a communication system comprising a BS with MAs and an IRS with movable elements was taken into account to remarkably enhance the system throughput. The authors of [28] proposed to mitigate double beam squint effect occurring in IRS-assisted wideband systems by embedding MAs and movable elements into the BS and IRS, respectively.

\begin{figure}[!t]
\centering
\includegraphics[scale=0.22]{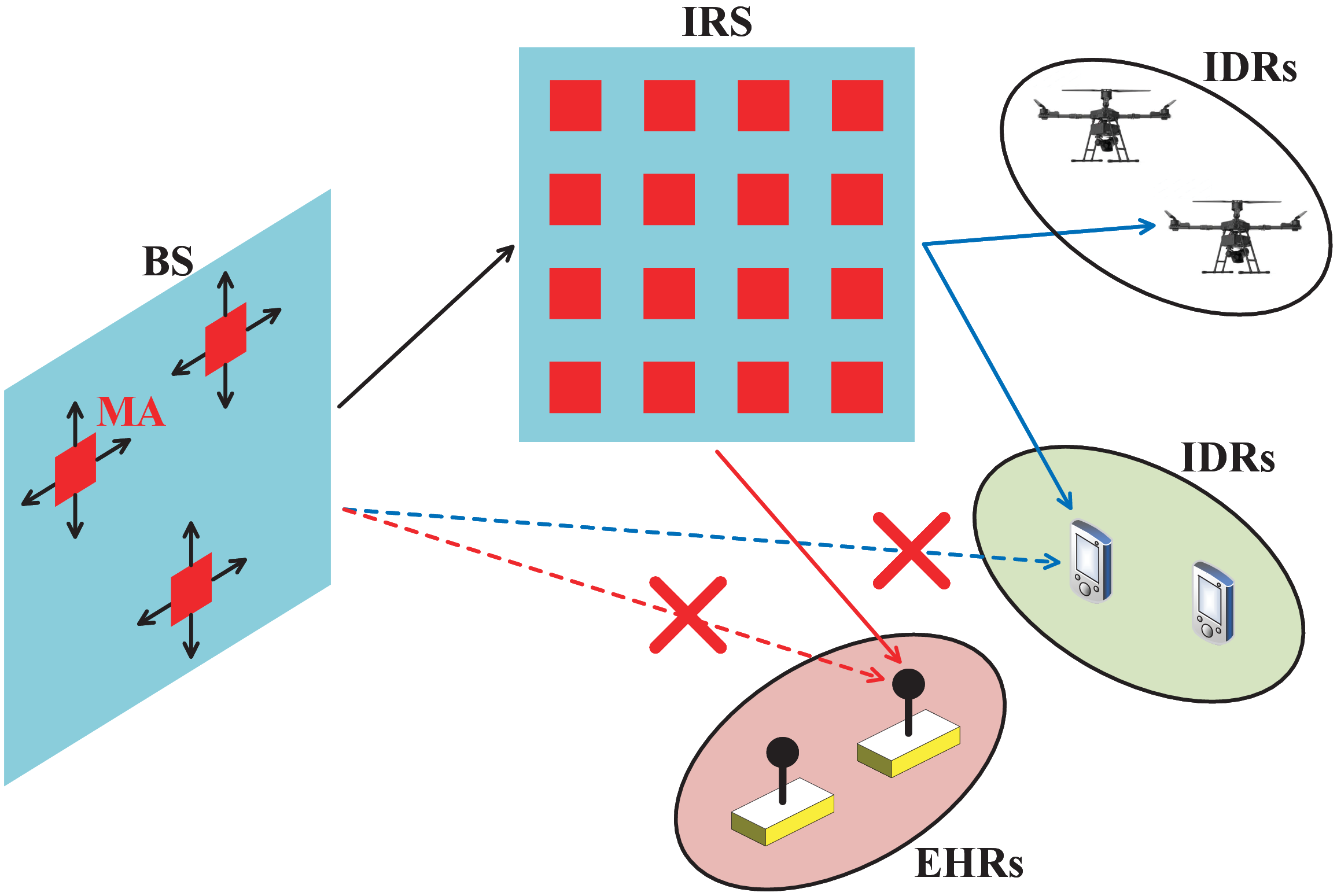}
\caption*{Fig. 1. IRS-aided SWIPT system with MAs.}
\end{figure}

Despite the aforementioned progresses that have been made in the existing literature, a thorough investigation for SWIPT systems empowered by both IRSs and MAs is still absent. Besides, although it has been demonstrated that SWIPT systems can dramatically benefit from IRSs or MAs [8]-[12], [20], [21], it still remains unknown that whether IRSs or MAs play a greater role in performance enhancement for SWIPT systems with both in a cooperative manner. These unresolved issues motivate this work, whose contributions are summarized as follows:

\begin{itemize}

\item To begin with, we propose a novel SWIPT system consisting of an MA-enabled BS and an IRS to promote data transmission for IDRs and power transfer for EHRs, as shown in Fig. 1. Via appropriately configuring MAs’ positions and IRS phase shifts, the capacity of considered SWIPT system can be significantly improved. Existing works have only studied SWIPT systems with IRSs or MAs being deployed [8]-[12], [20], [21]. To the best of authors’ knowledge, this is the first work taking an SWIPT system assisted by both MAs and an IRS into account.

\item Secondly, based on our proposed SWIPT system, we intend to maximize the weighted sum-rate of IDRs by jointly optimizing the BS active beamforming, the IRS passive beamforming, and MA configuration, while guaranteeing minimal receive power of all EHRs. This task is challenging since the associated optimization is nonconvex. To solve this problem, we first utilize the weighted minimal mean square error (WMMSE) [29] method to transform the original problem into a more tractable one, which is then tackled by block coordinate descent (BCD) framework [30]. Specifically, majorization-minimization (MM) method [31] is exploited to update BS beamforming and MAs’ positions, while cutting-the-edge penalty duality decomposition framework [32] is leveraged to adjust the phase shifts of IRS elements.

\item Thirdly, our formulated sum-rate maximization problem may be infeasible if the parameters are inappropriate, for example, the BS power budget is small while the requirement of each EHR is rather excessive. Hence, we develop a criterion to characterize its feasibility. To this end, we formulate an optimization problem for feasibility characterization and design an algorithm to solve it. Note that the feasibility characterization for quality-of-service (QoS) thresholds of devices such as IDRs and EHRs in SWIPT systems is highly meaningful, which, however, generally lacks investigation in the existing literature, e.g., [8], [9], [11], [12], [20], [21].

\item Last but not least, we present extensive numerical results to verify the effectiveness of our proposed solutions to sum-rate maximization problem and feasibility characterization, and the advantages of our proposed with the deployment of an IRS and MAs. Especially, we observe an interesting phenomenon that, under our considered scenario, the performance improvement yielded by MAs is less than that obtained from optimizing IRS phase shifts compared to the scheme with FPAs and random IRS configuration.

\end{itemize}

\section{System Model and Problem Formulation}

As shown in Fig. 1, we consider an IRS-aided SWIPT system, where a BS equipped with $M$ MAs serves $K_{\mathrm{I}}$ single-FPA IDRs and $K_{\mathrm{E}}$ single-FPA EHRs, and an IRS with $N$ reflecting elements is deployed in this system to enhance data transmission and power transfer. Without loss of generality, we assume that the BS is a 2-D planar array with a square shape of size $A \times A$. Denote $\mathcal{M} \triangleq \{ 1, \hdots, M \}$, $\mathcal{K}_{\mathrm{I}} \triangleq \{ 1, \hdots, K_{\mathrm{I}} \}$, $\mathcal{K}_{\mathrm{E}} \triangleq \{ 1, \hdots, K_{\mathrm{E}} \}$, and $\mathcal{N} \triangleq \{ 1, \hdots, N \}$ as the sets of MAs, IDRs, EHRs, and IRS elements, respectively. Besides, we define $\mathbf{t}_{\mathrm{B},m}, \; m \in \mathcal{M}$, as the relative position of the $m$-th MA w.r.t. the center of the BS array. Since all MAs should move within the BS surface, each MA should satisfy the constraint $\mathbf{t}_{\mathrm{B},m} \in \mathcal{C}_{\mathrm{B}} \triangleq \{ (x, y)|x \in [-\frac{A}{2}, \frac{A}{2}], y \in [-\frac{A}{2}, \frac{A}{2}] \}, \; m \in \mathcal{M}$, where $\mathcal{C}_{\mathrm{B}}$ stands for the feasible moving region of MAs [15], [16].

\subsection{Channel Models}

In this paper, we adopt quasi-static far-field channel models$^{1}$\footnote{$^{1}$In this paper, perfect channel state information (CSI) is assumed to be available, which can be obtained by various channel estimation methods [33]-[36].} [15], [16]. Besides, the direct links between the BS and IDRs/EHRs are assumed to be severely blocked [23], [24]. Based on these assumptions, we describe the channels of considered system. Taking the BS-IRS channel, defined as $\mathbf{G}$, into account first, we denote $L_{\mathrm{G}}^{\mathrm{t}}$ and $L_{\mathrm{G}}^{\mathrm{r}}$ as the number of $\mathbf{G}$'s transmit paths and receive paths, respectively. Furthermore, define $\vartheta_{\mathrm{G},l}^{\mathrm{t}}, \; l \in \{ 1, \hdots, L_{\mathrm{G}}^{\mathrm{t}} \}$, and $\varphi_{\mathrm{G},l}^{\mathrm{t}}, \; l \in \{ 1, \hdots, L_{\mathrm{G}}^{\mathrm{t}} \}$, as the azimuth angle-of-departure (AoD) and elevation AoD of the $l$-th transmit path, respectively. Then, the field response vector of the $m$-th MA is given by [15]
\begin{align}
\mathbf{f}_{\mathrm{G}}^{\mathrm{t}}(\mathbf{t}_{\mathrm{B},m}) \!=\! [e^{\jmath\frac{2\pi}{\lambda_{\mathrm{c}}}\mathbf{t}_{\mathrm{B},m}^{\mathsf{T}}\bm{\rho}_{\mathrm{G},1}^{\mathrm{t}}}, \hdots, e^{\jmath\frac{2\pi}{\lambda_{\mathrm{c}}}\mathbf{t}_{\mathrm{B},m}^{\mathsf{T}}\bm{\rho}_{\mathrm{G},L_{\mathrm{G}}^{\mathrm{t}}}^{\mathrm{t}}}]^{\mathsf{T}}, \; m \!\in\! \mathcal{M},
\end{align}
where $\bm{\rho}_{\mathrm{G},l}^{\mathrm{t}} \triangleq [\mathsf{cos}\vartheta_{\mathrm{G},l}^{\mathrm{t}}\mathsf{sin}\varphi_{\mathrm{G},l}^{\mathrm{t}}, \mathsf{cos}\varphi_{\mathrm{G},l}^{\mathrm{t}}]^{\mathsf{T}}, \; l \in \{ 1, \hdots, L_{\mathrm{G}}^{\mathrm{t}} \}$, and $\lambda_{\mathrm{c}}$ represents carrier wavelength. Via stacking the field response vectors of all MAs in a column-wise manner, the field response matrix of the BS array can be expressed as
\begin{align}
\mathbf{F}_{\mathrm{G}}^{\mathrm{t}}(\tilde{\mathbf{t}}_{\mathrm{B}}) = [\mathbf{f}_{\mathrm{G}}^{\mathrm{t}}(\mathbf{t}_{\mathrm{B},1}), \hdots, \mathbf{f}_{\mathrm{G}}^{\mathrm{t}}(\mathbf{t}_{\mathrm{B},M})],
\end{align}
where $\tilde{\mathbf{t}}_{\mathrm{B}} \triangleq [\mathbf{t}_{\mathrm{B},1}^{\mathsf{T}}, \hdots, \mathbf{t}_{\mathrm{B},M}^{\mathsf{T}}]^{\mathsf{T}}$ collects the positions of MAs. Similarly, denote $\vartheta_{\mathrm{G},l}^{\mathrm{r}}, \; l \in \{ 1, \hdots, L_{\mathrm{G}}^{\mathrm{r}} \}$, and $\varphi_{\mathrm{G},l}^{\mathrm{r}}, \; l \in \{ 1, \hdots, L_{\mathrm{G}}^{\mathrm{r}} \}$, as the azimuth angle-of-arrival (AoA) and elevation AoA of the $l$-th receive path, respectively. Then, the receive field response vector of the $n$-th IRS element is given as follows
\begin{align}
\mathbf{f}_{\mathrm{G}}^{\mathrm{r}}(\mathbf{t}_{\mathrm{R},n}) = [e^{\jmath\frac{2\pi}{\lambda_{\mathrm{c}}}\mathbf{t}_{\mathrm{R},n}^{\mathsf{T}}\bm{\rho}_{\mathrm{G},1}^{\mathrm{r}}}, \hdots, e^{\jmath\frac{2\pi}{\lambda_{\mathrm{c}}}\mathbf{t}_{\mathrm{R},n}^{\mathsf{T}}\bm{\rho}_{\mathrm{G},L_{\mathrm{G}}^{\mathrm{r}}}^{\mathrm{r}}}]^{\mathsf{T}}, \; n \in \mathcal{N},
\end{align}
where $\bm{\rho}_{\mathrm{G},l}^{\mathrm{r}} \triangleq [\mathsf{sin}\vartheta_{\mathrm{G},l}^{\mathrm{r}}\mathsf{sin}\varphi_{\mathrm{G},l}^{\mathrm{r}}, \mathsf{cos}\varphi_{\mathrm{G},l}^{\mathrm{r}}]^{\mathsf{T}}, \; l \in \{ 1, \hdots, L_{\mathrm{G}}^{\mathrm{r}} \}$, and $\mathbf{t}_{\mathrm{R},n}, \; n \in \mathcal{N}$, represents the relative position of the $n$-th IRS element w.r.t. the center of the IRS surface. By stacking the receive field response vectors of all reflecting elements column-by-column, the receive field response matrix of the IRS surface can be written as
\begin{align}
\mathbf{F}_{\mathrm{G}}^{\mathrm{r}}(\tilde{\mathbf{t}}_{\mathrm{R}}) = [\mathbf{f}_{\mathrm{G}}^{\mathrm{r}}(\mathbf{t}_{\mathrm{R},1}), \hdots, \mathbf{f}_{\mathrm{G}}^{\mathrm{r}}(\mathbf{t}_{\mathrm{R},N})],
\end{align}
where $\tilde{\mathbf{t}}_{\mathrm{R}} \triangleq [\mathbf{t}_{\mathrm{R},1}^{\mathsf{T}}, \hdots, \mathbf{t}_{\mathrm{R},N}^{\mathsf{T}}]^{\mathsf{T}}$ collects the positions of IRS elements. Based on the field response matrices $\mathbf{F}_{\mathrm{G}}^{\mathrm{t}}(\tilde{\mathbf{t}}_{\mathrm{B}})$ and $\mathbf{F}_{\mathrm{G}}^{\mathrm{r}}(\tilde{\mathbf{t}}_{\mathrm{R}})$, the BS-IRS channel $\mathbf{G}$ is given by
\begin{align}
\mathbf{G}(\tilde{\mathbf{t}}_{\mathrm{B}}) = (\mathbf{F}_{\mathrm{G}}^{\mathrm{r}}(\tilde{\mathbf{t}}_{\mathrm{R}}))^{\mathsf{H}}\bm{\Sigma}_{\mathrm{G}}\mathbf{F}_{\mathrm{G}}^{\mathrm{t}}(\tilde{\mathbf{t}}_{\mathrm{B}}),
\end{align}
where $\bm{\Sigma}_{\mathrm{G}}$ is the path response matrix from the center of the BS array to that of the IRS surface [15].

Next, we consider the channels between the IRS and IDRs. Define $L_{\mathrm{h},i}^{\mathrm{t}}, \; i \in \mathcal{K}_{\mathrm{I}}$, and $L_{\mathrm{h},i}^{\mathrm{r}}, \; i \in \mathcal{K}_{\mathrm{I}}$, as the number of transmit paths and receive paths between the IRS and the $i$-th IDR, respectively. Furthermore, we denote $\vartheta_{\mathrm{h},i,l}^{\mathrm{t}}, \; i \in \mathcal{K}_{\mathrm{I}}, \; l \in \{ 1, \hdots, L_{\mathrm{h},i}^{\mathrm{t}} \}$, and $\varphi_{\mathrm{h},i,l}^{\mathrm{t}}, \; i \in \mathcal{K}_{\mathrm{I}}, \; l \in \{ 1, \hdots, L_{\mathrm{h},i}^{\mathrm{t}} \}$, as the azimuth AoD and elevation AoD of the $l$-th transmit path between the IRS and the $i$-th IDR, respectively. Then, for the channel between the IRS and the $i$-th IDR, the transmit field response vector of the $n$-th IRS element is given by
\begin{align}
\mathbf{f}_{\mathrm{h},i}^{\mathrm{t}}(\mathbf{t}_{\mathrm{R},n}) &= [e^{\jmath\frac{2\pi}{\lambda_{\mathrm{c}}}\mathbf{t}_{\mathrm{R},n}^{\mathsf{T}}\bm{\rho}_{\mathrm{h},i,1}^{\mathrm{t}}}, \hdots, e^{\jmath\frac{2\pi}{\lambda_{\mathrm{c}}}\mathbf{t}_{\mathrm{R},n}^{\mathsf{T}}\bm{\rho}_{\mathrm{h},i,L_{\mathrm{h},i}^{\mathrm{t}}}^{\mathrm{t}}}]^{\mathsf{T}}, \notag\\
&\qquad\qquad\qquad\qquad\quad i \in \mathcal{K}_{\mathrm{I}}, \; n \in \mathcal{N},
\end{align}
where $\bm{\rho}_{\mathrm{h},i,l}^{\mathrm{t}} \triangleq [\mathsf{sin}\vartheta_{\mathrm{h},i,l}^{\mathrm{t}}\mathsf{sin}\varphi_{\mathrm{h},i,l}^{\mathrm{t}}, \mathsf{cos}\varphi_{\mathrm{h},i,l}^{\mathrm{t}}]^{\mathsf{T}}, \; i \in \mathcal{K}_{\mathrm{I}}, \; l \in \{ 1, \hdots, L_{\mathrm{h},i}^{\mathrm{t}} \}$. According to (6), the transmit field response matrix of the IRS surface for the channel between the IRS and the $i$-th IDR can be casted as
\begin{align}
\mathbf{F}_{\mathrm{h},i}^{\mathrm{t}}(\tilde{\mathbf{t}}_{\mathrm{R}}) = [\mathbf{f}_{\mathrm{h},i}^{\mathrm{t}}(\mathbf{t}_{\mathrm{R},1}), \hdots, \mathbf{f}_{\mathrm{h},i}^{\mathrm{t}}(\mathbf{t}_{\mathrm{R},N})], \; i \in \mathcal{K}_{\mathrm{I}},
\end{align}
and hence, the channel between the IRS and the $i$-th IDR reads
\begin{align}
\mathbf{h}_{\mathrm{r},i} = (\mathbf{F}_{\mathrm{h},i}^{\mathrm{t}}(\tilde{\mathbf{t}}_{\mathrm{R}}))^{\mathsf{H}}\bm{\Sigma}_{\mathrm{h},i}\bm{1}_{L_{\mathrm{h},i}^{\mathrm{r}}}, \; i \in \mathcal{K}_{\mathrm{I}},
\end{align}
where $\bm{\Sigma}_{\mathrm{h},i}, \; i \in \mathcal{K}_{\mathrm{I}}$, stands for the path response matrix from the center of the IRS surface to the $i$-th IDR.

Following similar procedure, the channel between the IRS and the $j$-th EHR can be expressed as
\begin{align}
\mathbf{g}_{\mathrm{r},j} = (\mathbf{F}_{\mathrm{g},j}^{\mathrm{t}}(\tilde{\mathbf{t}}_{\mathrm{R}}))^{\mathsf{H}}\bm{\Sigma}_{\mathrm{g},j}\bm{1}_{L_{\mathrm{g},j}^{\mathrm{r}}}, \; j \in \mathcal{K}_{\mathrm{E}},
\end{align}
where
\begin{align}
\mathbf{F}_{\mathrm{g},j}^{\mathrm{t}}(\tilde{\mathbf{t}}_{\mathrm{R}}) &= [\mathbf{f}_{\mathrm{g},j}^{\mathrm{t}}(\mathbf{t}_{\mathrm{R},1}), \hdots, \mathbf{f}_{\mathrm{g},j}^{\mathrm{t}}(\mathbf{t}_{\mathrm{R},N})], \; j \in \mathcal{K}_{\mathrm{E}}, \\
\mathbf{f}_{\mathrm{g},j}^{\mathrm{t}}(\mathbf{t}_{\mathrm{R},n}) &= [e^{\jmath\frac{2\pi}{\lambda_{\mathrm{c}}}\mathbf{t}_{\mathrm{R},n}^{\mathsf{T}}\bm{\rho}_{\mathrm{g},j,1}^{\mathrm{t}}}, \hdots, e^{\jmath\frac{2\pi}{\lambda_{\mathrm{c}}}\mathbf{t}_{\mathrm{R},n}^{\mathsf{T}}\bm{\rho}_{\mathrm{g},j,L_{\mathrm{g},j}^{\mathrm{t}}}^{\mathrm{t}}}]^{\mathsf{T}}, \notag\\
&\qquad\qquad\qquad\qquad\quad j \in \mathcal{K}_{\mathrm{E}}, \; n \in \mathcal{N},
\end{align}
with $\bm{\rho}_{\mathrm{g},j,l}^{\mathrm{t}} \triangleq [\mathsf{sin}\vartheta_{\mathrm{g},j,l}^{\mathrm{t}}\mathsf{sin}\varphi_{\mathrm{g},j,l}^{\mathrm{t}}, \mathsf{cos}\varphi_{\mathrm{g},j,l}^{\mathrm{t}}]^{\mathsf{T}}, \; j \in \mathcal{K}_{\mathrm{E}}, \; l \in \{ 1, \hdots, L_{\mathrm{g},j}^{\mathrm{t}} \}$. The parameters without description are similarly defined.

\subsection{Signal Models}

The signal transmitted from the BS is given by
\begin{align}
\mathbf{x} = \sum_{i=1}^{K_{\mathrm{I}}}\mathbf{f}_{\mathrm{I},i}s_{\mathrm{I},i} + \sum_{j=1}^{K_{\mathrm{E}}}\mathbf{f}_{\mathrm{E},j}s_{\mathrm{E},j},
\end{align}
where $\mathbf{f}_{\mathrm{I},i}, \; i \in \mathcal{K}_{\mathrm{I}}$, and $\mathbf{f}_{\mathrm{E},j}, \; j \in \mathcal{K}_{\mathrm{E}}$, denote the beamforming vectors for the $i$-th IDR and the $j$-th EHR, respectively, $s_{\mathrm{I},i}, \; i \in \mathcal{K}_{\mathrm{I}}$, and $s_{\mathrm{E},j}, \; j \in \mathcal{K}_{\mathrm{E}}$, represent the information-bearing and energy-carrying signals for the $i$-th IDR and the $j$-th EHR, respectively, which satisfy $s_{\mathrm{I},i} \sim \mathcal{CN}(0, 1), \; i \in \mathcal{K}_{\mathrm{I}}$, and $s_{\mathrm{E},j} \sim \mathcal{CN}(0, 1), \; j \in \mathcal{K}_{\mathrm{E}}$. For simplicity, we assume that $\mathbb{E} \{ s_{\mathrm{I},i}s_{\mathrm{I},k}^{*} \} = 0, \; i, k \in \mathcal{K}_{\mathrm{I}}, \; i \neq k$, $\mathbb{E} \{ s_{\mathrm{E},j}s_{\mathrm{E},k}^{*} \} = 0, \; j, k \in \mathcal{K}_{\mathrm{E}}, \; j \neq k$, and $\mathbb{E} \{ s_{\mathrm{I},i}s_{\mathrm{E},j}^{*} \} = 0, \; i \in \mathcal{K}_{\mathrm{I}}, \; j \in \mathcal{K}_{\mathrm{E}}$ [8], [9].

Define $\bm{\theta} \triangleq [e^{\jmath\theta_{1}}, \hdots, e^{\jmath\theta_{N}}]^{\mathsf{T}}$ as the IRS phase shift vector, where $\theta_{n}, \; n \in \mathcal{N}$, stands for the phase shift of the $n$-th reflecting element. Then, the signal received by the $i$-th IDR can be expressed as
\begin{align}
y_{\mathrm{I},i} = \mathbf{h}_{\mathrm{r},i}^{\mathsf{H}}\bm{\Theta}\mathbf{G}(\tilde{\mathbf{t}}_{\mathrm{B}})\mathbf{x} + n_{\mathrm{I},i}, \; i \in \mathcal{K}_{\mathrm{I}},
\end{align}
where $\bm{\Theta} \triangleq \mathsf{Diag}(\bm{\theta})$ and $n_{\mathrm{I},i}, \; i \in \mathcal{K}_{\mathrm{I}}$, is the thermal noise at the $i$-th IDR following the distribution $n_{\mathrm{I},i} \sim \mathcal{CN}(0, \sigma_{\mathrm{I},i}^{2})$, which is uncorrelated to transmit signals. Supposing that the interference caused by energy signals cannot be mitigated by IDRs [8], the SINR of the $i$-th IDR is given as follows
\begin{align}
\mathsf{SINR}_{i} &= |\mathbf{h}_{i}^{\mathsf{H}}(\tilde{\mathbf{t}}_{\mathrm{B}}, \bm{\Theta})\mathbf{f}_{\mathrm{I},i}|^{2}\bigg(\sum_{k=1, k \neq i}^{K_{\mathrm{I}}}|\mathbf{h}_{i}^{\mathsf{H}}(\tilde{\mathbf{t}}_{\mathrm{B}}, \bm{\Theta})\mathbf{f}_{\mathrm{I},k}|^{2} \notag\\
& + \sum_{j=1}^{K_{\mathrm{E}}}|\mathbf{h}_{i}^{\mathsf{H}}(\tilde{\mathbf{t}}_{\mathrm{B}}, \bm{\Theta})\mathbf{f}_{\mathrm{E},j}|^{2} + \sigma_{\mathrm{I},i}^{2}\bigg)^{-1}, \; i \in \mathcal{K}_{\mathrm{I}},
\end{align}
where $\mathbf{h}_{i}^{\mathsf{H}}(\tilde{\mathbf{t}}_{\mathrm{B}}, \bm{\Theta}) \triangleq \mathbf{h}_{\mathrm{r},i}^{\mathsf{H}}\bm{\Theta}\mathbf{G}(\tilde{\mathbf{t}}_{\mathrm{B}}), \; i \in \mathcal{K}_{\mathrm{I}}$, represents the equivalent channel between the BS and the $i$-th IDR.

On the other hand, the signal received by the $j$-th EHR can be written as
\begin{align}
y_{\mathrm{E},j} = \mathbf{g}_{\mathrm{r},j}^{\mathsf{H}}\bm{\Theta}\mathbf{G}(\tilde{\mathbf{t}}_{\mathrm{B}})\mathbf{x} + n_{\mathrm{E},j}, \; j \in \mathcal{K}_{\mathrm{E}},
\end{align}
where $n_{\mathrm{E},j}, \; j \in \mathcal{K}_{\mathrm{E}}$, is the thermal noise at the $j$-th EHR. By ignoring the noise power at the EHRs, the received RF power at the $j$-th EHR can be calculated as
\begin{align}
\mathsf{P}_{j} \!\!=\!\! \sum_{k=1}^{K_{\mathrm{I}}}|\mathbf{g}_{j}^{\mathsf{H}}(\tilde{\mathbf{t}}_{\mathrm{B}}, \bm{\Theta})\mathbf{f}_{\mathrm{I},k}|^{2} \!\!+\!\! \sum_{k=1}^{K_{\mathrm{E}}}|\mathbf{g}_{j}^{\mathsf{H}}(\tilde{\mathbf{t}}_{\mathrm{B}}, \bm{\Theta})\mathbf{f}_{\mathrm{E},k}|^{2}, j \!\in\! \mathcal{K}_{\mathrm{E}},
\end{align}
where $\mathbf{g}_{j}^{\mathsf{H}}(\tilde{\mathbf{t}}_{\mathrm{B}}, \bm{\Theta}) \triangleq \mathbf{g}_{\mathrm{r},j}^{\mathsf{H}}\bm{\Theta}\mathbf{G}(\tilde{\mathbf{t}}_{\mathrm{B}}), \; j \in \mathcal{K}_{\mathrm{E}}$, stands for the equivalent channel between the BS and the $j$-th EHR.

\subsection{Problem Formulation}

In this paper, we aim at maximizing the weighted sum-rate of IDRs by jointly optimizing the beamforming at the BS, the positions of MAs, and the configuration of IRS, while guaranteeing the power harvest at each EHRs. Its optimization problem can be formulated as
\begin{align}
(\mathcal{P}1): \max_{\tilde{\mathbf{f}}, \tilde{\mathbf{t}}_{\mathrm{B}}, \bm{\theta}} \; &\sum_{i=1}^{K_{\mathrm{I}}}\alpha_{i}\log (1 + \mathsf{SINR}_{i}) \\
\mathrm{s.t.} \; &\mathsf{P}_{j} \geq P_{\mathrm{E},j}, \; j \in \mathcal{K}_{\mathrm{E}}, \tag{17a}\\
&\sum_{i=1}^{K_{\mathrm{I}}}\| \mathbf{f}_{\mathrm{I},i} \|_{2}^{2} + \sum_{j=1}^{K_{\mathrm{E}}}\| \mathbf{f}_{\mathrm{E},j} \|_{2}^{2} \leq P_{\mathrm{B}}, \tag{17b}\\
&0 \leq \theta_{n} \leq 2\pi, \; n \in \mathcal{N}, \tag{17c}\\
&\mathbf{t}_{\mathrm{B},m} \in \mathcal{C}_{\mathrm{B}}, \; m \in \mathcal{M}, \tag{17d}\\
&\| \mathbf{t}_{\mathrm{B},m} \!-\! \mathbf{t}_{\mathrm{B},s} \|_{2} \!\geq\! D_{\mathrm{B}}, \; m, s \!\in\! \mathcal{M}, \; m \!\neq\! s, \tag{17e}
\end{align}
where $\tilde{\mathbf{f}} \triangleq [\mathbf{f}_{\mathrm{I},1}^{\mathsf{T}}, \hdots, \mathbf{f}_{\mathrm{I},K_{\mathrm{I}}}^{\mathsf{T}}, \mathbf{f}_{\mathrm{E},1}^{\mathsf{T}}, \hdots, \mathbf{f}_{\mathrm{E},K_{\mathrm{E}}}^{\mathsf{T}}]^{\mathsf{T}}$ collects the beamforming vectors for the IDRs and EHRs, and $\alpha_{i} \geq 0, \; i \in \mathcal{K}_{\mathrm{I}}$, denotes the rate weight of the $i$-th IDR. Constraint (17a) guarantees that the received RF power at the $j$-th EHR should exceed the minimal power harvest $P_{\mathrm{E},j}$, constraint (17b) describes that the total transmit power at the BS should lower than the power budget $P_{\mathrm{B}}$, constraint (17c) states the feasible phase shift of each IRS elements, constraints (17d) and (17e) characterize that each MA should move within the feasible region $\mathcal{C}_{\mathrm{B}}$, while keeping a distance of at least $D_{\mathrm{B}}$ from other MAs to avoid antenna coupling [16], [17]. Due to the logarithm form of objective function and non-convexity of constraints (17a), (17c), and (17e), $(\mathcal{P}1)$ is challenging to solve. In the following, we propose an algorithm to tackle $(\mathcal{P}1)$.

\section{Proposed Algorithm}

In this section, we develop an algorithm to handle $(\mathcal{P}1)$. Specifically, we propose to utilize WMMSE method [29] to equivalently transform $(\mathcal{P}1)$ into a more tractable optimization, which is then solved based on BCD framework [30]. Note in the following, we use $\mathbf{F}_{\mathrm{G}}^{\mathrm{r}}$, $\mathbf{G}$, $\mathbf{h}_{i}, \; i \in \mathcal{K}_{\mathrm{I}}$, and $\mathbf{g}_{j}, \; j \in \mathcal{K}_{\mathrm{E}}$, to respectively replace $\mathbf{F}_{\mathrm{G}}^{\mathrm{r}}(\tilde{\mathbf{t}}_{\mathrm{R}})$, $\mathbf{G}(\tilde{\mathbf{t}}_{\mathrm{B}})$, $\mathbf{h}_{i}(\tilde{\mathbf{t}}_{\mathrm{B}}, \bm{\Theta}), \; i \in \mathcal{K}_{\mathrm{I}}$, and $\mathbf{g}_{j}(\tilde{\mathbf{t}}_{\mathrm{B}}, \bm{\Theta}), \; j \in \mathcal{K}_{\mathrm{E}}$, for brevity. Besides, it is assumed that the feasible domain of $(\mathcal{P}1)$ is non-empty, which can be examined by our proposed feasibility characterization method (see Sec. IV).

\subsection{WMMSE Transformation}

\begin{figure*}[!t]
\normalsize
\setcounter{MYtempeqncnt}{\value{equation}}
\setcounter{equation}{17}
\begin{align}
\log (1 + \mathsf{SINR}_{i}) &\geq \log (w_{i}) - w_{i}\bigg(1 - |\mathbf{h}_{i}^{\mathsf{H}}\mathbf{f}_{\mathrm{I},i}|^{2}\bigg(\sum_{k=1}^{K_{\mathrm{I}}}|\mathbf{h}_{i}^{\mathsf{H}}\mathbf{f}_{\mathrm{I},k}|^{2} + \sum_{j=1}^{K_{\mathrm{E}}}|\mathbf{h}_{i}^{\mathsf{H}}\mathbf{f}_{\mathrm{E},j}|^{2} + \sigma_{\mathrm{I},i}^{2}\bigg)^{-1}\bigg) + 1 \notag\\
& \geq \log (w_{i}) - w_{i}\bigg(|v_{i}|^{2}\bigg(\sum_{k=1}^{K_{\mathrm{I}}}|\mathbf{h}_{i}^{\mathsf{H}}\mathbf{f}_{\mathrm{I},k}|^{2} + \sum_{j=1}^{K_{\mathrm{E}}}|\mathbf{h}_{i}^{\mathsf{H}}\mathbf{f}_{\mathrm{E},j}|^{2} + \sigma_{\mathrm{I},i}^{2}\bigg) - 2\mathsf{Re} \{ v_{i}^{*}\mathbf{h}_{i}^{\mathsf{H}}\mathbf{f}_{\mathrm{I},i} \} + 1\bigg) + 1 \triangleq \mathsf{R}_{i}, \; i \in \mathcal{K}_{\mathrm{I}}.
\end{align}
\setcounter{equation}{\value{MYtempeqncnt}}
\hrulefill
\vspace*{4pt}
\end{figure*}

\setcounter{equation}{18}

To begin with, it can be observed that the rate objective function (17) makes $(\mathcal{P}1)$ difficult to solve. To overcome this difficulty, we exploit WMMSE approach [29] to equivalently transform (17) into a more tractable form. Specifically, by introducing auxiliary variables $\mathbf{v} = [v_{1}, \hdots, v_{K_{\mathrm{I}}}]^{\mathsf{T}}$ and $\mathbf{w} = [w_{1}, \hdots, w_{K_{\mathrm{I}}}]^{\mathsf{T}}$, the inequalities in (18) hold according to WMMSE method, which is shown at the top of the next page. Then, $(\mathcal{P}1)$ can be equivalently recasted as
\begin{align}
(\mathcal{P}2): \max_{\substack{\mathbf{v}, \mathbf{w}, \\ \tilde{\mathbf{f}}, \tilde{\mathbf{t}}_{\mathrm{B}}, \bm{\theta}}} \; &\sum_{i=1}^{K_{\mathrm{I}}}\alpha_{i}\mathsf{R}_{i} \\
\mathrm{s.t.} \; &\mathsf{P}_{j} \geq P_{\mathrm{E},j}, \; j \in \mathcal{K}_{\mathrm{E}}, \tag{19a}\\
&\sum_{i=1}^{K_{\mathrm{I}}}\| \mathbf{f}_{\mathrm{I},i} \|_{2}^{2} + \sum_{j=1}^{K_{\mathrm{E}}}\| \mathbf{f}_{\mathrm{E},j} \|_{2}^{2} \leq P_{\mathrm{B}}, \tag{19b}\\
&0 \leq \theta_{n} \leq 2\pi, \; n \in \mathcal{N}, \tag{19c}\\
&\mathbf{t}_{\mathrm{B},m} \in \mathcal{C}_{\mathrm{B}}, \; m \in \mathcal{M}, \tag{19d}\\
&\| \mathbf{t}_{\mathrm{B},m} \!-\! \mathbf{t}_{\mathrm{B},s} \|_{2} \!\geq\! D_{\mathrm{B}}, \; m, s \!\in\! \mathcal{M}, \; m \!\neq\! s, \tag{19e}
\end{align}
and we proceed to tackle $(\mathcal{P}2)$ in the sequel.

\subsection{Solution to $(\mathcal{P}2)$}

In the following, we propose to leverage BCD framework [30] to tackle $(\mathcal{P}2)$ by iteratively updating one variable at a time with other variables being fixed.

\subsubsection{Updating Auxiliary Variable $\mathbf{v}$} When other variables are given, the subproblem w.r.t. $\mathbf{v}$ is given by
\begin{align}
\min_{v_{i}} \; &|v_{i}|^{2}\bigg(\sum_{k=1}^{K_{\mathrm{I}}}|\mathbf{h}_{i}^{\mathsf{H}}\mathbf{f}_{\mathrm{I},k}|^{2} + \sum_{j=1}^{K_{\mathrm{E}}}|\mathbf{h}_{i}^{\mathsf{H}}\mathbf{f}_{\mathrm{E},j}|^{2} + \sigma_{\mathrm{I},i}^{2}\bigg) \notag\\
& - 2\mathsf{Re} \{ v_{i}^{*}\mathbf{h}_{i}^{\mathsf{H}}\mathbf{f}_{\mathrm{I},i} \}, \; i \in \mathcal{K}_{\mathrm{I}},
\end{align}
which is an unconstrained convex quadratic optimization problem. Therefore, by checking the first-order optimality condition, the optimal solution to $\mathbf{v}$ can be expressed as
\begin{align}
v_{i} = \frac{\mathbf{h}_{i}^{\mathsf{H}}\mathbf{f}_{\mathrm{I},i}}{\sum_{k=1}^{K_{\mathrm{I}}}|\mathbf{h}_{i}^{\mathsf{H}}\mathbf{f}_{\mathrm{I},k}|^{2} + \sum_{j=1}^{K_{\mathrm{E}}}|\mathbf{h}_{i}^{\mathsf{H}}\mathbf{f}_{\mathrm{E},j}|^{2} + \sigma_{\mathrm{I},i}^{2}}, \; i \in \mathcal{K}_{\mathrm{I}}.
\end{align}

\subsubsection{Updating Auxiliary Variable $\mathbf{w}$} With other variables being fixed, the subproblem w.r.t. $\mathbf{w}$ can be written as
\begin{align}
\max_{w_{i}} \; &\log (w_{i}) \!-\! w_{i}\bigg(|v_{i}|^{2}\bigg(\sum_{k=1}^{K_{\mathrm{I}}}|\mathbf{h}_{i}^{\mathsf{H}}\mathbf{f}_{\mathrm{I},k}|^{2} \!+\! \sum_{j=1}^{K_{\mathrm{E}}}|\mathbf{h}_{i}^{\mathsf{H}}\mathbf{f}_{\mathrm{E},j}|^{2} \!+\! \sigma_{\mathrm{I},i}^{2}\bigg) \notag\\
& - 2\mathsf{Re} \{ v_{i}^{*}\mathbf{h}_{i}^{\mathsf{H}}\mathbf{f}_{\mathrm{I},i} \} + 1\bigg), \; i \in \mathcal{K}_{\mathrm{I}}.
\end{align}
This is an unconstrained convex optimization problem, whose optimal solution can be obtained by setting the first-order derivative of objective function w.r.t. $w_{i}$ to zero, given as
\begin{align}
w_{i} &= \bigg(|v_{i}|^{2}\bigg(\sum_{k=1}^{K_{\mathrm{I}}}|\mathbf{h}_{i}^{\mathsf{H}}\mathbf{f}_{\mathrm{I},k}|^{2} + \sum_{j=1}^{K_{\mathrm{E}}}|\mathbf{h}_{i}^{\mathsf{H}}\mathbf{f}_{\mathrm{E},j}|^{2} + \sigma_{\mathrm{I},i}^{2}\bigg) \notag\\
& - 2\mathsf{Re} \{ v_{i}^{*}\mathbf{h}_{i}^{\mathsf{H}}\mathbf{f}_{\mathrm{I},i} \} + 1\bigg)^{-1}, \; i \in \mathcal{K}_{\mathrm{I}}.
\end{align}

\subsubsection{Updating BS Beamforming $\tilde{\mathbf{f}}$} Next, we optimize the BS beamforming vectors for IDRs and EHRs, which implies to solve the following optimization
\begin{align}
\min_{\tilde{\mathbf{f}}} \; &\sum_{i=1}^{K_{\mathrm{I}}}\alpha_{i}w_{i}\bigg(|v_{i}|^{2}\bigg(\sum_{k=1}^{K_{\mathrm{I}}}|\mathbf{h}_{i}^{\mathsf{H}}\mathbf{f}_{\mathrm{I},k}|^{2} + \sum_{j=1}^{K_{\mathrm{E}}}|\mathbf{h}_{i}^{\mathsf{H}}\mathbf{f}_{\mathrm{E},j}|^{2}\bigg) \notag\\
& - 2\mathsf{Re} \{ v_{i}^{*}\mathbf{h}_{i}^{\mathsf{H}}\mathbf{f}_{\mathrm{I},i} \} \bigg) \\
\mathrm{s.t.} \; &\mathsf{P}_{j} \geq P_{\mathrm{E},j}, \; j \in \mathcal{K}_{\mathrm{E}}, \tag{24a}\\
&\sum_{i=1}^{K_{\mathrm{I}}}\| \mathbf{f}_{\mathrm{I},i} \|_{2}^{2} + \sum_{j=1}^{K_{\mathrm{E}}}\| \mathbf{f}_{\mathrm{E},j} \|_{2}^{2} \leq P_{\mathrm{B}}, \tag{24b}
\end{align}
which is still challenging due to the nonconvex constraint (24a). To tackle this issue, we utilize MM methodology [31] to convexify (24a). Noticing that the left hand side of (24a) is convex w.r.t. $\tilde{\mathbf{f}}$, we can acquire a concave lower bound surrogate by its first-order Taylor expansion. Specifically, the first-order Taylor expansion of $\mathsf{P}_{j}$ w.r.t. $\tilde{\mathbf{f}}$ at the point $\tilde{\mathbf{f}}^{(n)}$ reads
\begin{align}
\mathsf{P}_{j} &\geq \sum_{k=1}^{K_{\mathrm{I}}}\bigg(2\mathsf{Re} \{ (\mathbf{f}_{\mathrm{I},k}^{(n)})^{\mathsf{H}}\mathbf{g}_{j}\mathbf{g}_{j}^{\mathsf{H}}\mathbf{f}_{\mathrm{I},k} \} - |\mathbf{g}_{j}^{\mathsf{H}}(\mathbf{f}_{\mathrm{I},k}^{(n)})|^{2}\bigg) \notag\\
& + \sum_{k=1}^{K_{\mathrm{E}}}\bigg(2\mathsf{Re} \{ (\mathbf{f}_{\mathrm{E},k}^{(n)})^{\mathsf{H}}\mathbf{g}_{j}\mathbf{g}_{j}^{\mathsf{H}}\mathbf{f}_{\mathrm{E},k} \} - |\mathbf{g}_{j}^{\mathsf{H}}(\mathbf{f}_{\mathrm{E},k}^{(n)})|^{2}\bigg) \notag\\
& \triangleq \mathsf{P}_{j}(\tilde{\mathbf{f}}|\tilde{\mathbf{f}}^{(n)}), \; j \in \mathcal{K}_{\mathrm{E}}.
\end{align}
After substituting (25) into (24a), we obtain the following convex optimization problem
\begin{align}
(\mathcal{P}3): \min_{\tilde{\mathbf{f}}} \; &\sum_{i=1}^{K_{\mathrm{I}}}\alpha_{i}w_{i}\bigg(|v_{i}|^{2}\bigg(\sum_{k=1}^{K_{\mathrm{I}}}|\mathbf{h}_{i}^{\mathsf{H}}\mathbf{f}_{\mathrm{I},k}|^{2} + \sum_{j=1}^{K_{\mathrm{E}}}|\mathbf{h}_{i}^{\mathsf{H}}\mathbf{f}_{\mathrm{E},j}|^{2}\bigg) \notag\\
& - 2\mathsf{Re} \{ v_{i}^{*}\mathbf{h}_{i}^{\mathsf{H}}\mathbf{f}_{\mathrm{I},i} \} \bigg) \\
\mathrm{s.t.} \; &\mathsf{P}_{j}(\tilde{\mathbf{f}}|\tilde{\mathbf{f}}^{(n)}) \geq P_{\mathrm{E},j}, \; j \in \mathcal{K}_{\mathrm{E}}, \tag{26a}\\
&\sum_{i=1}^{K_{\mathrm{I}}}\| \mathbf{f}_{\mathrm{I},i} \|_{2}^{2} + \sum_{j=1}^{K_{\mathrm{E}}}\| \mathbf{f}_{\mathrm{E},j} \|_{2}^{2} \leq P_{\mathrm{B}}, \tag{26b}
\end{align}
which can be solved by numerical solvers, e.g., CVX [37].

\subsubsection{Updating IRS Configuration $\bm{\theta}$} In the following, we consider the optimization for updating $\bm{\theta}$ when other variables are fixed, whose compact form can be casted as follows after some manipulations
\begin{align}
\min_{\bm{\theta}} \; &\bm{\theta}^{\mathsf{H}}\mathbf{Q}_{0}\bm{\theta} - 2\mathsf{Re} \{ \mathbf{q}_{0}^{\mathsf{H}}\bm{\theta} \} \\
\mathrm{s.t.} \; &\bm{\theta}^{\mathsf{H}}\mathbf{Q}_{1,j}\bm{\theta} \geq P_{\mathrm{E},j}, \; j \in \mathcal{K}_{\mathrm{E}}, \tag{27a}\\
&0 \leq \theta_{n} \leq 2\pi, \; n \in \mathcal{N}, \tag{27b}
\end{align}
where
\begin{align}
\mathbf{Q}_{0} &\triangleq \sum_{i=1}^{K_{\mathrm{I}}}\alpha_{i}w_{i}|v_{i}|^{2}\bigg(\sum_{k=1}^{K_{\mathrm{I}}}\mathsf{Diag}^{*}(\mathbf{G}\mathbf{f}_{\mathrm{I},k})\mathbf{h}_{\mathrm{r},i}\mathbf{h}_{\mathrm{r},i}^{\mathsf{H}}\mathsf{Diag}(\mathbf{G}\mathbf{f}_{\mathrm{I},k}) \notag\\
& + \sum_{j=1}^{K_{\mathrm{E}}}\mathsf{Diag}^{*}(\mathbf{G}\mathbf{f}_{\mathrm{E},j})\mathbf{h}_{\mathrm{r},i}\mathbf{h}_{\mathrm{r},i}^{\mathsf{H}}\mathsf{Diag}(\mathbf{G}\mathbf{f}_{\mathrm{E},j})\bigg), \notag\\
\mathbf{q}_{0}^{\mathsf{H}} &\triangleq \sum_{i=1}^{K_{\mathrm{I}}}\alpha_{i}w_{i}v_{i}^{*}\mathbf{h}_{\mathrm{r},i}^{\mathsf{H}}\mathsf{Diag}(\mathbf{G}\mathbf{f}_{\mathrm{I},i}), \notag\\
\mathbf{Q}_{1,j} &\triangleq \sum_{k=1}^{K_{\mathrm{I}}}\mathsf{Diag}^{*}(\mathbf{G}\mathbf{f}_{\mathrm{I},k})\mathbf{g}_{\mathrm{r},j}\mathbf{g}_{\mathrm{r},j}^{\mathsf{H}}\mathsf{Diag}(\mathbf{G}\mathbf{f}_{\mathrm{I},k}) \notag\\
& + \sum_{k=1}^{K_{\mathrm{E}}}\mathsf{Diag}^{*}(\mathbf{G}\mathbf{f}_{\mathrm{E},k})\mathbf{g}_{\mathrm{r},j}\mathbf{g}_{\mathrm{r},j}^{\mathsf{H}}\mathsf{Diag}(\mathbf{G}\mathbf{f}_{\mathrm{E},k}), \; j \in \mathcal{K}_{\mathrm{E}}.
\end{align}
The above optimization problem is nonconvex because constraint (27a) is nonconvex and constraint (27b) has a constant modulus form. To attack this problem, we propose to exploit PDD methodology [32]. Specifically, via introducing an auxiliary variable $\bm{\phi}$, the above problem can be rewritten as
\begin{align}
\min_{\bm{\theta}, \bm{\phi}} \; &\bm{\theta}^{\mathsf{H}}\mathbf{Q}_{0}\bm{\theta} - 2\mathsf{Re} \{ \mathbf{q}_{0}^{\mathsf{H}}\bm{\theta} \} \\
\mathrm{s.t.} \; &\bm{\theta}^{\mathsf{H}}\mathbf{Q}_{1,j}\bm{\theta} \geq P_{\mathrm{E},j}, \; j \in \mathcal{K}_{\mathrm{E}}, \tag{29a}\\
&\bm{\theta} = \bm{\phi}, \tag{29b}\\
&|[\bm{\theta}]_{n}| \leq 1, \; n \in \mathcal{N}, \tag{29c}\\
&|[\bm{\phi}]_{n}| = 1, \; n \in \mathcal{N}. \tag{29d}
\end{align}
Based on PDD framework [32], we penalize the equality constraint (29b) and turn to tackle its augmented Lagrangian problem, which is given by
\begin{align}
\min_{\bm{\theta}, \bm{\phi}} \; &\bm{\theta}^{\mathsf{H}}\mathbf{Q}_{0}\bm{\theta} - 2\mathsf{Re} \{ \mathbf{q}_{0}^{\mathsf{H}}\bm{\theta} \} + \frac{1}{2\rho}\| \bm{\theta} - \bm{\phi} + \rho\bm{\lambda} \|_{2}^{2} \\
\mathrm{s.t.} \; &\bm{\theta}^{\mathsf{H}}\mathbf{Q}_{1,j}\bm{\theta} \geq P_{\mathrm{E},j}, \; j \in \mathcal{K}_{\mathrm{E}}, \tag{30a}\\
&|[\bm{\theta}]_{n}| \leq 1, \; n \in \mathcal{N}, \tag{30b}\\
&|[\bm{\phi}]_{n}| = 1, \; n \in \mathcal{N}, \tag{30c}
\end{align}
where $\rho$ and $\bm{\lambda}$ represent the penalty coefficient and Lagrangian multiplier, respectively.

The PDD method comprises a two-layer procedure [32], with its inner layer alternatively optimizing the variables $\bm{\theta}$ and $\bm{\phi}$, and its outer layer updating the Lagrangian multiplier $\bm{\lambda}$ or penalty coefficient $\rho$.

Taking the inner layer into account, we leverage the block successive upper-bound minimization (BSUM) framework [32], [38] to update $\bm{\theta}$ and $\bm{\phi}$. With $\bm{\phi}$ being fixed, the subproblem w.r.t. $\bm{\theta}$ can be expressed as
\begin{align}
\min_{\bm{\theta}} \; &\bm{\theta}^{\mathsf{H}}\mathbf{Q}_{0}\bm{\theta} - 2\mathsf{Re} \{ \mathbf{q}_{0}^{\mathsf{H}}\bm{\theta} \} + \frac{1}{2\rho}\| \bm{\theta} - \bm{\phi} + \rho\bm{\lambda} \|_{2}^{2} \\
\mathrm{s.t.} \; &\bm{\theta}^{\mathsf{H}}\mathbf{Q}_{1,j}\bm{\theta} \geq P_{\mathrm{E},j}, \; j \in \mathcal{K}_{\mathrm{E}}, \tag{31a}\\
&|[\bm{\theta}]_{n}| \leq 1, \; n \in \mathcal{N}, \tag{31b}
\end{align}
which is still difficult to solve due to the nonconvex constraint (31a). We again adopt MM approach [31] to handle (31a). It can be observed that the left hand side of (31a) is convex w.r.t. $\bm{\theta}$, therefore, we utilize its first-order Taylor expansion at the point $\bm{\theta}^{(n)}$ to construct a concave lower bound, which is given as follows
\begin{align}
\bm{\theta}^{\mathsf{H}}\mathbf{Q}_{1,j}\bm{\theta} &\geq 2\mathsf{Re} \{ (\bm{\theta}^{(n)})^{\mathsf{H}}\mathbf{Q}_{1,j}\bm{\theta} \} - (\bm{\theta}^{(n)})^{\mathsf{H}}\mathbf{Q}_{1,j}\bm{\theta}^{(n)} \notag\\
& \triangleq \mathsf{P}_{j}(\bm{\theta}|\bm{\theta}^{(n)}), \; j \in \mathcal{K}_{\mathrm{E}}.
\end{align}
After replacing the left hand side of (31a) by (32), the following optimization problem can be obtained
\begin{align}
(\mathcal{P}4): \min_{\bm{\theta}} \; &\bm{\theta}^{\mathsf{H}}\mathbf{Q}_{0}\bm{\theta} - 2\mathsf{Re} \{ \mathbf{q}_{0}^{\mathsf{H}}\bm{\theta} \} + \frac{1}{2\rho}\| \bm{\theta} - \bm{\phi} + \rho\bm{\lambda} \|_{2}^{2} \\
\mathrm{s.t.} \; &\mathsf{P}_{j}(\bm{\theta}|\bm{\theta}^{(n)}) \geq P_{\mathrm{E},j}, \; j \in \mathcal{K}_{\mathrm{E}}, \tag{33a}\\
&|[\bm{\theta}]_{n}| \leq 1, \; n \in \mathcal{N}. \tag{33b}
\end{align}
Problem $(\mathcal{P}4)$ is convex, which can be numerically solved.

\begin{algorithm}[!t]
\caption{PDD-Based Solution to $\bm{\theta}$}
\begin{algorithmic}[1]
\STATE Initialize feasible $\bm{\theta}^{(0)}$, $\bm{\phi}^{(0)}$, $\bm{\lambda}^{(0)}$, $\rho^{(0)}$, and $n = 0$;
\REPEAT
\STATE set $\bm{\theta}^{(n, 0)} := \bm{\theta}^{(n)}$, $\bm{\phi}^{(n, 0)} := \bm{\phi}^{(n)}$, and $i = 0$;
\REPEAT
\STATE update $\bm{\theta}^{(n, i + 1)}$ by solving $(\mathcal{P}4)$;
\STATE update $\bm{\phi}^{(n, i + 1)}$ by (36);
\STATE set $i := i + 1$;
\UNTIL{convergence}
\STATE set $\bm{\theta}^{(n + 1)} := \bm{\theta}^{(n, \infty)}$ and $\bm{\phi}^{(n + 1)} := \bm{\phi}^{(n, \infty)}$;
\IF{$\| \bm{\theta}^{(n + 1)} - \bm{\phi}^{(n + 1)} \|_{\infty} \leq \delta$}
\STATE $\bm{\lambda}^{(n + 1)} := \bm{\lambda}^{(n)} + \frac{1}{\rho^{(n)}}(\bm{\theta}^{(n + 1)} - \bm{\phi}^{(n + 1)})$, $\rho^{(n + 1)} := \rho^{(n)}$;
\ELSE
\STATE $\bm{\lambda}^{(n + 1)} := \bm{\lambda}^{(n)}$, $\rho^{(n + 1)} := c_{\rho} \cdot \rho^{(n)}$;
\ENDIF
\STATE set $n := n + 1$;
\UNTIL{$\| \bm{\theta}^{(n)} - \bm{\phi}^{(n)} \|_{2}$ is sufficiently small}
\end{algorithmic}
\end{algorithm}

For another, when $\bm{\theta}$ is given, the update of $\bm{\phi}$ is given by
\begin{align}
\min_{\bm{\phi}} \; &\| \bm{\theta} - \bm{\phi} + \rho\bm{\lambda} \|_{2}^{2} \\
\mathrm{s.t.} \; &|[\bm{\phi}]_{n}| = 1, \; n \in \mathcal{N}. \tag{34a}
\end{align}
Note that the objective function (34) can be further derived as $\| \bm{\theta} - \bm{\phi} + \rho\bm{\lambda} \|_{2}^{2} = \| \bm{\phi} \|_{2}^{2} - 2\mathsf{Re} \{ (\bm{\theta} + \rho\bm{\lambda})^{\mathsf{H}}\bm{\phi} \} + \| \bm{\theta} + \rho\bm{\lambda} \|_{2}^{2} \overset{(a)}{=} N - 2\mathsf{Re} \{ (\bm{\theta} + \rho\bm{\lambda})^{\mathsf{H}}\bm{\phi} \} + \| \bm{\theta} + \rho\bm{\lambda} \|_{2}^{2}$, where (a) follows the constraint (34a). Consequently, via omitting the constant terms unrelated to $\bm{\phi}$, the above problem reduces to
\begin{align}
\max_{\bm{\phi}} \; &\mathsf{Re} \{ (\bm{\theta} + \rho\bm{\lambda})^{\mathsf{H}}\bm{\phi} \} \\
\mathrm{s.t.} \; &|[\bm{\phi}]_{n}| = 1, \; n \in \mathcal{N}, \tag{35a}
\end{align}
which is still nonconvex resulting from the constant modulus constraint (35a). Nevertheless, this problem admits a closed-form solution, written as
\begin{align}
\bm{\phi} = e^{\jmath\angle(\bm{\theta} + \rho\bm{\lambda})}.
\end{align}

Considering the outer layer, if the equality constraint (29b) approximately holds, we update the Lagrangian multiplier $\bm{\lambda}$ as $\bm{\lambda} := \bm{\lambda} + \rho^{-1}(\bm{\theta} - \bm{\phi})$. Otherwise, the penalty coefficient $\rho$ needs to be decreased such that $\bm{\theta} = \bm{\phi}$ is forced to be approached.

The proposed PDD algorithm for updating $\bm{\theta}$ is summarized in Algorithm 1, where $c_{\rho} \in (0, 1)$ is used to adjust $\rho$.

\subsubsection{Updating MAs' Positions $\tilde{\mathbf{t}}_{\mathrm{B}}$} Finally, we study the configuration of MAs. To acquire a more tractable solution, merely one MA's position is updated at a time with other variables being given. Specifically, the subproblem w.r.t. the position of the $m$-th MA, for $m \in \mathcal{M}$, can be expressed as
\begin{align}
\min_{\mathbf{t}_{\mathrm{B},m}} \; &\sum_{i=1}^{K_{\mathrm{I}}}\alpha_{i}w_{i}\bigg(|v_{i}|^{2}\bigg(\sum_{k=1}^{K_{\mathrm{I}}}|\mathbf{h}_{i}^{\mathsf{H}}\mathbf{f}_{\mathrm{I},k}|^{2} + \sum_{j=1}^{K_{\mathrm{E}}}|\mathbf{h}_{i}^{\mathsf{H}}\mathbf{f}_{\mathrm{E},j}|^{2}\bigg) \notag\\
& - 2\mathsf{Re} \{ v_{i}^{*}\mathbf{h}_{i}^{\mathsf{H}}\mathbf{f}_{\mathrm{I},i} \} \bigg) \\
\mathrm{s.t.} \; &\mathsf{P}_{j} \geq P_{\mathrm{E},j}, \; j \in \mathcal{K}_{\mathrm{E}}, \tag{37a}\\
&\mathbf{t}_{\mathrm{B},m} \in \mathcal{C}_{\mathrm{B}}, \tag{37b}\\
&\| \mathbf{t}_{\mathrm{B},m} - \mathbf{t}_{\mathrm{B},s} \|_{2} \geq D_{\mathrm{B}}, \; s \in \mathcal{M}, \; m \neq s, \tag{37c}
\end{align}
which is nonconvex since (37), (37a), and (37c) are nonconvex w.r.t. $\mathbf{t}_{\mathrm{B},m}$. To tackle this issue, we invoke MM method [31] to explore a convex surrogate for (37), and concave lower bounds for the left hand side of (37a) and that of (37c), respectively, which are elaborated in the following.

Firstly, we handle the non-convexity caused by (37). To this end, denote $\tilde{\mathsf{R}}(\mathbf{t}_{\mathrm{B},m})$ as the objective function (37), which can be equivalently recasted as
\begin{align}
\tilde{\mathsf{R}}(\mathbf{t}_{\mathrm{B},m}) &= (\mathbf{f}_{\mathrm{G}}^{\mathrm{t}}(\mathbf{t}_{\mathrm{B},m}))^{\mathsf{H}}\mathbf{R}_{0,m}\mathbf{f}_{\mathrm{G}}^{\mathrm{t}}(\mathbf{t}_{\mathrm{B},m}) \notag\\
& - 2\mathsf{Re} \{ \mathbf{r}_{0,m}^{\mathsf{H}}\mathbf{f}_{\mathrm{G}}^{\mathrm{t}}(\mathbf{t}_{\mathrm{B},m}) \} + r_{0,m},
\end{align}
where
\begin{align}
&\mathbf{R}_{0,m} \triangleq \sum_{i=1}^{K_{\mathrm{I}}}\alpha_{i}w_{i}|v_{i}|^{2}\bigg(\sum_{k=1}^{K_{\mathrm{I}}}(\mathbf{h}_{\mathrm{r},i}^{\mathsf{H}}\bm{\Theta}(\mathbf{F}_{\mathrm{G}}^{\mathrm{r}})^{\mathsf{H}}\bm{\Sigma}_{\mathrm{G}}[\mathbf{f}_{\mathrm{I},k}]_{m})^{\mathsf{H}} \notag\\
& \times \mathbf{h}_{\mathrm{r},i}^{\mathsf{H}}\bm{\Theta}(\mathbf{F}_{\mathrm{G}}^{\mathrm{r}})^{\mathsf{H}}\bm{\Sigma}_{\mathrm{G}}[\mathbf{f}_{\mathrm{I},k}]_{m} + \sum_{j=1}^{K_{\mathrm{E}}}(\mathbf{h}_{\mathrm{r},i}^{\mathsf{H}}\bm{\Theta}(\mathbf{F}_{\mathrm{G}}^{\mathrm{r}})^{\mathsf{H}}\bm{\Sigma}_{\mathrm{G}}[\mathbf{f}_{\mathrm{E},j}]_{m})^{\mathsf{H}} \notag\\
& \times \mathbf{h}_{\mathrm{r},i}^{\mathsf{H}}\bm{\Theta}(\mathbf{F}_{\mathrm{G}}^{\mathrm{r}})^{\mathsf{H}}\bm{\Sigma}_{\mathrm{G}}[\mathbf{f}_{\mathrm{E},j}]_{m}\bigg), \notag\\
&\mathbf{r}_{0,m}^{\mathsf{H}} \triangleq \sum_{i=1}^{K_{\mathrm{I}}}\alpha_{i}w_{i}\bigg(v_{i}^{*}\mathbf{h}_{\mathrm{r},i}^{\mathsf{H}}\bm{\Theta}(\mathbf{F}_{\mathrm{G}}^{\mathrm{r}})^{\mathsf{H}}\bm{\Sigma}_{\mathrm{G}}[\mathbf{f}_{\mathrm{I},i}]_{m} \notag\\
&\qquad - \sum_{k=1}^{K_{\mathrm{I}}}|v_{i}|^{2}C_{1,i,k,m}^{*}\mathbf{h}_{\mathrm{r},i}^{\mathsf{H}}\bm{\Theta}(\mathbf{F}_{\mathrm{G}}^{\mathrm{r}})^{\mathsf{H}}\bm{\Sigma}_{\mathrm{G}}[\mathbf{f}_{\mathrm{I},k}]_{m} \notag\\
&\qquad - \sum_{j=1}^{K_{\mathrm{E}}}|v_{i}|^{2}C_{2,i,j,m}^{*}\mathbf{h}_{\mathrm{r},i}^{\mathsf{H}}\bm{\Theta}(\mathbf{F}_{\mathrm{G}}^{\mathrm{r}})^{\mathsf{H}}\bm{\Sigma}_{\mathrm{G}}[\mathbf{f}_{\mathrm{E},j}]_{m}\bigg), \notag\\
&r_{0,m} \triangleq \sum_{i=1}^{K_{\mathrm{I}}}\alpha_{i}w_{i}\bigg(|v_{i}|^{2}\bigg(\sum_{k=1}^{K_{\mathrm{I}}}|C_{1,i,k,m}|^{2} + \sum_{j=1}^{K_{\mathrm{E}}}|C_{2,i,j,m}|^{2}\bigg) \notag\\
&\qquad - 2\mathsf{Re} \{ v_{i}^{*}C_{1,i,i,m} \} \bigg),
\end{align}
with (for $i \in \mathcal{K}_{\mathrm{I}}$ and $j \in \mathcal{K}_{\mathrm{E}}$)
\begin{align}
C_{1,i,k,m} &\triangleq \sum_{n=1, n \neq m}^{M}\mathbf{h}_{\mathrm{r},i}^{\mathsf{H}}\bm{\Theta}(\mathbf{F}_{\mathrm{G}}^{\mathrm{r}})^{\mathsf{H}}\bm{\Sigma}_{\mathrm{G}}\mathbf{f}_{\mathrm{G}}^{\mathrm{t}}(\mathbf{t}_{\mathrm{B},n})[\mathbf{f}_{\mathrm{I},k}]_{n}, \notag\\
C_{2,i,j,m} &\triangleq \sum_{n=1, n \neq m}^{M}\mathbf{h}_{\mathrm{r},i}^{\mathsf{H}}\bm{\Theta}(\mathbf{F}_{\mathrm{G}}^{\mathrm{r}})^{\mathsf{H}}\bm{\Sigma}_{\mathrm{G}}\mathbf{f}_{\mathrm{G}}^{\mathrm{t}}(\mathbf{t}_{\mathrm{B},n})[\mathbf{f}_{\mathrm{E},j}]_{n}.
\end{align}
It can be observed from (38) that $\tilde{\mathsf{R}}(\mathbf{t}_{\mathrm{B},m})$ is convex w.r.t. $\mathbf{f}_{\mathrm{G}}^{\mathrm{t}}(\mathbf{t}_{\mathrm{B},m})$. Hence, according to MM framework [31], an upper bound of $\tilde{\mathsf{R}}(\mathbf{t}_{\mathrm{B},m})$ can be obtained by utilizing the second-order Taylor expansion of $(\mathbf{f}_{\mathrm{G}}^{\mathrm{t}}(\mathbf{t}_{\mathrm{B},m}))^{\mathsf{H}}\mathbf{R}_{0,m}\mathbf{f}_{\mathrm{G}}^{\mathrm{t}}(\mathbf{t}_{\mathrm{B},m})$ with its Hessian matrix being surrogated by a larger one. Specifically, the second-order Taylor expansion of $(\mathbf{f}_{\mathrm{G}}^{\mathrm{t}}(\mathbf{t}_{\mathrm{B},m}))^{\mathsf{H}}\mathbf{R}_{0,m}\mathbf{f}_{\mathrm{G}}^{\mathrm{t}}(\mathbf{t}_{\mathrm{B},m})$ w.r.t. $\mathbf{f}_{\mathrm{G}}^{\mathrm{t}}(\mathbf{t}_{\mathrm{B},m})$ at the point $\mathbf{f}_{\mathrm{G}}^{\mathrm{t}}(\mathbf{t}_{\mathrm{B},m}^{(n)})$ is given by
\begin{align}
&(\mathbf{f}_{\mathrm{G}}^{\mathrm{t}}(\mathbf{t}_{\mathrm{B},m}))^{\mathsf{H}}\mathbf{R}_{0,m}\mathbf{f}_{\mathrm{G}}^{\mathrm{t}}(\mathbf{t}_{\mathrm{B},m}) = (\mathbf{f}_{\mathrm{G}}^{\mathrm{t}}(\mathbf{t}_{\mathrm{B},m}^{(n)}))^{\mathsf{H}}\mathbf{R}_{0,m}\mathbf{f}_{\mathrm{G}}^{\mathrm{t}}(\mathbf{t}_{\mathrm{B},m}^{(n)}) \notag\\
& + 2\mathsf{Re} \{ (\mathbf{f}_{\mathrm{G}}^{\mathrm{t}}(\mathbf{t}_{\mathrm{B},m}^{(n)}))^{\mathsf{H}}\mathbf{R}_{0,m}(\mathbf{f}_{\mathrm{G}}^{\mathrm{t}}(\mathbf{t}_{\mathrm{B},m}) - \mathbf{f}_{\mathrm{G}}^{\mathrm{t}}(\mathbf{t}_{\mathrm{B},m}^{(n)})) \} \notag\\
& + (\mathbf{f}_{\mathrm{G}}^{\mathrm{t}}(\mathbf{t}_{\mathrm{B},m}) \!-\! \mathbf{f}_{\mathrm{G}}^{\mathrm{t}}(\mathbf{t}_{\mathrm{B},m}^{(n)}))^{\mathsf{H}}\mathbf{R}_{0,m}(\mathbf{f}_{\mathrm{G}}^{\mathrm{t}}(\mathbf{t}_{\mathrm{B},m}) \!-\! \mathbf{f}_{\mathrm{G}}^{\mathrm{t}}(\mathbf{t}_{\mathrm{B},m}^{(n)})),
\end{align}
which can be upper-bounded by
\begin{align}
&(\mathbf{f}_{\mathrm{G}}^{\mathrm{t}}(\mathbf{t}_{\mathrm{B},m}))^{\mathsf{H}}\mathbf{R}_{0,m}\mathbf{f}_{\mathrm{G}}^{\mathrm{t}}(\mathbf{t}_{\mathrm{B},m}) \leq (\mathbf{f}_{\mathrm{G}}^{\mathrm{t}}(\mathbf{t}_{\mathrm{B},m}^{(n)}))^{\mathsf{H}}\mathbf{R}_{0,m}\mathbf{f}_{\mathrm{G}}^{\mathrm{t}}(\mathbf{t}_{\mathrm{B},m}^{(n)}) \notag\\
& + 2\mathsf{Re} \{ (\mathbf{f}_{\mathrm{G}}^{\mathrm{t}}(\mathbf{t}_{\mathrm{B},m}^{(n)}))^{\mathsf{H}}\mathbf{R}_{0,m}(\mathbf{f}_{\mathrm{G}}^{\mathrm{t}}(\mathbf{t}_{\mathrm{B},m}) - \mathbf{f}_{\mathrm{G}}^{\mathrm{t}}(\mathbf{t}_{\mathrm{B},m}^{(n)})) \} \notag\\
& + \lambda_{\mathrm{max}}(\mathbf{R}_{0,m})\| \mathbf{f}_{\mathrm{G}}^{\mathrm{t}}(\mathbf{t}_{\mathrm{B},m}) \!-\! \mathbf{f}_{\mathrm{G}}^{\mathrm{t}}(\mathbf{t}_{\mathrm{B},m}^{(n)}) \|_{2}^{2},
\end{align}
where $\lambda_{\mathrm{max}}(\mathbf{X})$ represents the maximal eigenvalue of $\mathbf{X}$. After substituting (42) into (38), we have
\begin{align}
&\tilde{\mathsf{R}}(\mathbf{t}_{\mathrm{B},m}) \leq 2\mathsf{Re} \{ \mathbf{b}_{0,m}^{\mathsf{H}}\mathbf{f}_{\mathrm{G}}^{\mathrm{t}}(\mathbf{t}_{\mathrm{B},m}) \} - (\mathbf{f}_{\mathrm{G}}^{\mathrm{t}}(\mathbf{t}_{\mathrm{B},m}^{(n)}))^{\mathsf{H}}\mathbf{R}_{0,m}\mathbf{f}_{\mathrm{G}}^{\mathrm{t}}(\mathbf{t}_{\mathrm{B},m}^{(n)}) \notag\\
& + \lambda_{\mathrm{max}}(\mathbf{R}_{0,m})(\| \mathbf{f}_{\mathrm{G}}^{\mathrm{t}}(\mathbf{t}_{\mathrm{B},m}) \|_{2}^{2} + \| \mathbf{f}_{\mathrm{G}}^{\mathrm{t}}(\mathbf{t}_{\mathrm{B},m}^{(n)}) \|_{2}^{2}) + r_{0,m} \notag\\
& \overset{(a)}{=} 2\mathsf{Re} \{ \mathbf{b}_{0,m}^{\mathsf{H}}\mathbf{f}_{\mathrm{G}}^{\mathrm{t}}(\mathbf{t}_{\mathrm{B},m}) \} - (\mathbf{f}_{\mathrm{G}}^{\mathrm{t}}(\mathbf{t}_{\mathrm{B},m}^{(n)}))^{\mathsf{H}}\mathbf{R}_{0,m}\mathbf{f}_{\mathrm{G}}^{\mathrm{t}}(\mathbf{t}_{\mathrm{B},m}^{(n)}) \notag\\
& + \lambda_{\mathrm{max}}(\mathbf{R}_{0,m})(L_{\mathrm{G}}^{\mathrm{t}} + \| \mathbf{f}_{\mathrm{G}}^{\mathrm{t}}(\mathbf{t}_{\mathrm{B},m}^{(n)}) \|_{2}^{2}) + r_{0,m},
\end{align}
where $\mathbf{b}_{0,m}^{\mathsf{H}} \triangleq (\mathbf{f}_{\mathrm{G}}^{\mathrm{t}}(\mathbf{t}_{\mathrm{B},m}^{(n)}))^{\mathsf{H}}(\mathbf{R}_{0,m} - \lambda_{\mathrm{max}}(\mathbf{R}_{0,m})\mathbf{I}_{L_{\mathrm{G}}^{\mathrm{t}}}) - \mathbf{r}_{0,m}^{\mathsf{H}}$, and (a) holds since the elements within $\mathbf{f}_{\mathrm{G}}^{\mathrm{t}}(\mathbf{t}_{\mathrm{B},m})$ are unit modulus. By dropping the constant terms independent of $\mathbf{t}_{\mathrm{B},m}$ in (43), we turn to tackle the newly acquired objective, which can be expressed as
\begin{align}
\bar{\mathsf{R}}(\mathbf{t}_{\mathrm{B},m}) = \mathsf{Re} \{ \mathbf{b}_{0,m}^{\mathsf{H}}\mathbf{f}_{\mathrm{G}}^{\mathrm{t}}(\mathbf{t}_{\mathrm{B},m}) \}.
\end{align}

However, $\bar{\mathsf{R}}(\mathbf{t}_{\mathrm{B},m})$ is still neither convex nor concave w.r.t. $\mathbf{t}_{\mathrm{B},m}$. To overcome this difficulty, we again exploit MM method [31] to construct a convex upper bound of $\bar{\mathsf{R}}(\mathbf{t}_{\mathrm{B},m})$ via examining its second-order Taylor expansion and replacing the Hessian matrix by a larger one. Specifically, $\bar{\mathsf{R}}(\mathbf{t}_{\mathrm{B},m})$ can be first derived as
\begin{align}
\bar{\mathsf{R}}(\mathbf{t}_{\mathrm{B},m}) &= \sum_{l=1}^{L_{\mathrm{G}}^{\mathrm{t}}}|[\mathbf{b}_{0,m}]_{l}|\mathsf{cos}\bigg(\frac{2\pi}{\lambda_{\mathrm{c}}}\mathbf{t}_{\mathrm{B},m}^{\mathsf{T}}\bm{\rho}_{\mathrm{G},l}^{\mathrm{t}} - \angle([\mathbf{b}_{0,m}]_{l})\bigg) \notag\\
& = \sum_{l=1}^{L_{\mathrm{G}}^{\mathrm{t}}}|[\mathbf{b}_{0,m}]_{l}|\mathsf{cos}(\kappa_{m,l}^{\mathrm{R}}),
\end{align}
where $\kappa_{m,l}^{\mathrm{R}} \triangleq \frac{2\pi}{\lambda_{\mathrm{c}}}\mathbf{t}_{\mathrm{B},m}^{\mathsf{T}}\bm{\rho}_{\mathrm{G},l}^{\mathrm{t}} - \angle([\mathbf{b}_{0,m}]_{l}), \; l \in \{ 1, \hdots, L_{\mathrm{G}}^{\mathrm{t}} \}$. Then, its second-order Taylor expansion w.r.t. $\mathbf{t}_{\mathrm{B},m}$ at the point $\mathbf{t}_{\mathrm{B},m}^{(n)}$ is given as follows
\begin{align}
\bar{\mathsf{R}}(\mathbf{t}_{\mathrm{B},m}) &\approx \nabla_{\mathbf{t}_{\mathrm{B},m}}^{\mathsf{T}}(\bar{\mathsf{R}}(\mathbf{t}_{\mathrm{B},m}))|_{\mathbf{t}_{\mathrm{B},m} = \mathbf{t}_{\mathrm{B},m}^{(n)}}(\mathbf{t}_{\mathrm{B},m} - \mathbf{t}_{\mathrm{B},m}^{(n)}) \notag\\
& + (\mathbf{t}_{\mathrm{B},m} - \mathbf{t}_{\mathrm{B},m}^{(n)})^{\mathsf{T}}\frac{\nabla_{\mathbf{t}_{\mathrm{B},m}}^{2}(\bar{\mathsf{R}}(\mathbf{t}_{\mathrm{B},m}))|_{\mathbf{t}_{\mathrm{B},m} = \mathbf{t}_{\mathrm{B},m}^{(n)}}}{2} \notag\\
& \times (\mathbf{t}_{\mathrm{B},m} - \mathbf{t}_{\mathrm{B},m}^{(n)}) + \bar{\mathsf{R}}(\mathbf{t}_{\mathrm{B},m}^{(n)}),
\end{align}
where the gradient $\nabla_{\mathbf{t}_{\mathrm{B},m}}(\bar{\mathsf{R}}(\mathbf{t}_{\mathrm{B},m}))$ and Hessian matrix $\nabla_{\mathbf{t}_{\mathrm{B},m}}^{2}(\bar{\mathsf{R}}(\mathbf{t}_{\mathrm{B},m}))$ are respectively written as
\begin{align}
&\nabla_{\mathbf{t}_{\mathrm{B},m}}(\bar{\mathsf{R}}(\mathbf{t}_{\mathrm{B},m})) = \bigg[\frac{\partial\bar{\mathsf{R}}(\mathbf{t}_{\mathrm{B},m})}{\partial[\mathbf{t}_{\mathrm{B},m}]_{1}}, \frac{\partial\bar{\mathsf{R}}(\mathbf{t}_{\mathrm{B},m})}{\partial[\mathbf{t}_{\mathrm{B},m}]_{2}}\bigg]^{\mathsf{T}}, \notag\\
&\frac{\partial\bar{\mathsf{R}}(\mathbf{t}_{\mathrm{B},m})}{\partial[\mathbf{t}_{\mathrm{B},m}]_{1}} = -\sum_{l=1}^{L_{\mathrm{G}}^{\mathrm{t}}}|[\mathbf{b}_{0,m}]_{l}|\frac{2\pi}{\lambda_{\mathrm{c}}}[\bm{\rho}_{\mathrm{G},l}^{\mathrm{t}}]_{1}\mathsf{sin}(\kappa_{m,l}^{\mathrm{R}}), \notag\\
&\frac{\partial\bar{\mathsf{R}}(\mathbf{t}_{\mathrm{B},m})}{\partial[\mathbf{t}_{\mathrm{B},m}]_{2}} = -\sum_{l=1}^{L_{\mathrm{G}}^{\mathrm{t}}}|[\mathbf{b}_{0,m}]_{l}|\frac{2\pi}{\lambda_{\mathrm{c}}}[\bm{\rho}_{\mathrm{G},l}^{\mathrm{t}}]_{2}\mathsf{sin}(\kappa_{m,l}^{\mathrm{R}}), \\
&\nabla_{\mathbf{t}_{\mathrm{B},m}}^{2}(\bar{\mathsf{R}}(\mathbf{t}_{\mathrm{B},m})) = \begin{bmatrix}
\frac{\partial^{2}\bar{\mathsf{R}}(\mathbf{t}_{\mathrm{B},m})}{\partial[\mathbf{t}_{\mathrm{B},m}]_{1}^{2}} & \frac{\partial^{2}\bar{\mathsf{R}}(\mathbf{t}_{\mathrm{B},m})}{\partial[\mathbf{t}_{\mathrm{B},m}]_{1}\partial[\mathbf{t}_{\mathrm{B},m}]_{2}} \\
\frac{\partial^{2}\bar{\mathsf{R}}(\mathbf{t}_{\mathrm{B},m})}{\partial[\mathbf{t}_{\mathrm{B},m}]_{2}\partial[\mathbf{t}_{\mathrm{B},m}]_{1}} & \frac{\partial^{2}\bar{\mathsf{R}}(\mathbf{t}_{\mathrm{B},m})}{\partial[\mathbf{t}_{\mathrm{B},m}]_{2}^{2}} \\
\end{bmatrix}, \notag\\
&\frac{\partial^{2}\bar{\mathsf{R}}(\mathbf{t}_{\mathrm{B},m})}{\partial[\mathbf{t}_{\mathrm{B},m}]_{1}^{2}} = -\sum_{l=1}^{L_{\mathrm{G}}^{\mathrm{t}}}|[\mathbf{b}_{0,m}]_{l}|\frac{4\pi^{2}}{\lambda_{\mathrm{c}}^{2}}[\bm{\rho}_{\mathrm{G},l}^{\mathrm{t}}]_{1}^{2}\mathsf{cos}(\kappa_{m,l}^{\mathrm{R}}), \notag\\
&\frac{\partial^{2}\bar{\mathsf{R}}(\mathbf{t}_{\mathrm{B},m})}{\partial[\mathbf{t}_{\mathrm{B},m}]_{1}\partial[\mathbf{t}_{\mathrm{B},m}]_{2}} \!\!=\!\! -\!\!\sum_{l=1}^{L_{\mathrm{G}}^{\mathrm{t}}}|[\mathbf{b}_{0,m}]_{l}|\frac{4\pi^{2}}{\lambda_{\mathrm{c}}^{2}}[\bm{\rho}_{\mathrm{G},l}^{\mathrm{t}}]_{1}[\bm{\rho}_{\mathrm{G},l}^{\mathrm{t}}]_{2}\mathsf{cos}(\kappa_{m,l}^{\mathrm{R}}), \notag\\
&\frac{\partial^{2}\bar{\mathsf{R}}(\mathbf{t}_{\mathrm{B},m})}{\partial[\mathbf{t}_{\mathrm{B},m}]_{2}\partial[\mathbf{t}_{\mathrm{B},m}]_{1}} = \frac{\partial^{2}\bar{\mathsf{R}}(\mathbf{t}_{\mathrm{B},m})}{\partial[\mathbf{t}_{\mathrm{B},m}]_{1}\partial[\mathbf{t}_{\mathrm{B},m}]_{2}}, \notag\\
&\frac{\partial^{2}\bar{\mathsf{R}}(\mathbf{t}_{\mathrm{B},m})}{\partial[\mathbf{t}_{\mathrm{B},m}]_{2}^{2}} = -\sum_{l=1}^{L_{\mathrm{G}}^{\mathrm{t}}}|[\mathbf{b}_{0,m}]_{l}|\frac{4\pi^{2}}{\lambda_{\mathrm{c}}^{2}}[\bm{\rho}_{\mathrm{G},l}^{\mathrm{t}}]_{2}^{2}\mathsf{cos}(\kappa_{m,l}^{\mathrm{R}}).
\end{align}
Furthermore, noticing that the curvature of $\nabla_{\mathbf{t}_{\mathrm{B},m}}^{2}(\bar{\mathsf{R}}(\mathbf{t}_{\mathrm{B},m}))$ is bounded, it can be concluded that there must exist a positive semidefinite matrix $\mathbf{M}_{m}^{\mathrm{R}}$ such that $\mathbf{M}_{m}^{\mathrm{R}} \succeq \nabla_{\mathbf{t}_{\mathrm{B},m}}^{2}(\bar{\mathsf{R}}(\mathbf{t}_{\mathrm{B},m}))$, whose derivation is relegated to Appendix A. By substituting $\mathbf{M}_{m}^{\mathrm{R}}$ into (46), a convex upper bound of $\bar{\mathsf{R}}(\mathbf{t}_{\mathrm{B},m})$ can be achieved. Finally, omitting the constant terms unrelated to $\mathbf{t}_{\mathrm{B},m}$, we obtain a convex objective w.r.t. $\mathbf{t}_{\mathrm{B},m}$, given by
\begin{align}
\tilde{\mathsf{R}}(\mathbf{t}_{\mathrm{B},m}|\mathbf{t}_{\mathrm{B},m}^{(n)}) &\!=\! \mathbf{t}_{\mathrm{B},m}^{\mathsf{T}}\frac{\mathbf{M}_{m}^{\mathrm{R}}}{2}\mathbf{t}_{\mathrm{B},m} \!+\! (\nabla_{\mathbf{t}_{\mathrm{B},m}}^{\mathsf{T}}(\bar{\mathsf{R}}(\mathbf{t}_{\mathrm{B},m}))|_{\mathbf{t}_{\mathrm{B},m} = \mathbf{t}_{\mathrm{B},m}^{(n)}} \notag\\
& \!-\! \mathbf{t}_{\mathrm{B},m}^{(n)}\mathbf{M}_{m}^{\mathrm{R}})\mathbf{t}_{\mathrm{B},m}.
\end{align}

Secondly, we proceed to convexify (37a). Note that the left hand side of (37a) is convex w.r.t. $\mathbf{g}_{j}$, hence, based on MM framework [31], its first-order Taylor expansion w.r.t. $\mathbf{g}_{j}$ at the point $\mathbf{g}_{j}^{(n)}$ serves as a lower bound, i.e.,
\begin{align}
\mathsf{P}_{j} &\geq 2\mathsf{Re} \{ (\mathbf{g}_{j}^{(n)})^{\mathsf{H}}\mathbf{F}\mathbf{g}_{j} \} - (\mathbf{g}_{j}^{(n)})^{\mathsf{H}}\mathbf{F}\mathbf{g}_{j}^{(n)} \notag\\
& = 2\mathsf{Re} \{ \mathbf{b}_{1,j,m}^{\mathsf{H}}\mathbf{f}_{\mathrm{G}}^{\mathrm{t}}(\mathbf{t}_{\mathrm{B},m}) \} + C_{3,j,m}, \; j \in \mathcal{K}_{\mathrm{E}},
\end{align}
where
\begin{align}
\mathbf{F} &\triangleq \sum_{k=1}^{K_{\mathrm{I}}}\mathbf{f}_{\mathrm{I},k}\mathbf{f}_{\mathrm{I},k}^{\mathsf{H}} + \sum_{k=1}^{K_{\mathrm{E}}}\mathbf{f}_{\mathrm{E},k}\mathbf{f}_{\mathrm{E},k}^{\mathsf{H}}, \notag\\
\mathbf{b}_{1,j,m}^{\mathsf{H}} &\triangleq \mathbf{g}_{\mathrm{r},j}^{\mathsf{H}}\bm{\Theta}(\mathbf{F}_{\mathrm{G}}^{\mathrm{r}})^{\mathsf{H}}\bm{\Sigma}_{\mathrm{G}}[\mathbf{F}\mathbf{g}_{j}^{(n)}]_{m}, \; j \in \mathcal{K}_{\mathrm{E}}, \notag\\
C_{3,j,m} &\triangleq 2\mathsf{Re} \bigg\{ \sum_{i=1, i \neq m}^{M}\mathbf{g}_{\mathrm{r},j}^{\mathsf{H}}\bm{\Theta}(\mathbf{F}_{\mathrm{G}}^{\mathrm{r}})^{\mathsf{H}}\bm{\Sigma}_{\mathrm{G}}\mathbf{f}_{\mathrm{G}}^{\mathrm{t}}(\mathbf{t}_{\mathrm{B},i})[\mathbf{F}\mathbf{g}_{j}^{(n)}]_{i} \bigg\} \notag\\
& - (\mathbf{g}_{j}^{(n)})^{\mathsf{H}}\mathbf{F}\mathbf{g}_{j}^{(n)}, \; j \in \mathcal{K}_{\mathrm{E}}.
\end{align}

Nevertheless, the right hand side of (50) is still neither convex nor concave w.r.t. $\mathbf{t}_{\mathrm{B},m}$. To tackle this issue, we again leverage MM approach [31] to construct a concave lower bound of $\bar{\mathsf{P}}_{j}(\mathbf{t}_{\mathrm{B},m}) \triangleq 2\mathsf{Re} \{ \mathbf{b}_{1,j,m}^{\mathsf{H}}\mathbf{f}_{\mathrm{G}}^{\mathrm{t}}(\mathbf{t}_{\mathrm{B},m}) \}, \; j \in \mathcal{K}_{\mathrm{E}}$, via using its second-order Taylor expansion with the Hessian matrix being replaced by a smaller one. Specifically, $\bar{\mathsf{P}}_{j}(\mathbf{t}_{\mathrm{B},m})$ can be first rewritten as
\begin{align}
\bar{\mathsf{P}}_{j}(\mathbf{t}_{\mathrm{B},m}) &= 2\sum_{l=1}^{L_{\mathrm{G}}^{\mathrm{t}}}|[\mathbf{b}_{1,j,m}]_{l}|\mathsf{cos}\bigg(\frac{2\pi}{\lambda_{\mathrm{c}}}\mathbf{t}_{\mathrm{B},m}^{\mathsf{T}}\bm{\rho}_{\mathrm{G},l}^{\mathrm{t}} - \angle([\mathbf{b}_{1,j,m}]_{l})\bigg) \notag\\
& = 2\sum_{l=1}^{L_{\mathrm{G}}^{\mathrm{t}}}|[\mathbf{b}_{1,j,m}]_{l}|\mathsf{cos}(\kappa_{j,m,l}^{\mathrm{P}}), \; j \in \mathcal{K}_{\mathrm{E}},
\end{align}
where $\kappa_{j,m,l}^{\mathrm{P}} \triangleq \frac{2\pi}{\lambda_{\mathrm{c}}}\mathbf{t}_{\mathrm{B},m}^{\mathsf{T}}\bm{\rho}_{\mathrm{G},l}^{\mathrm{t}} - \angle([\mathbf{b}_{1,j,m}]_{l}), \; j \in \mathcal{K}_{\mathrm{E}}, \; l \in \{ 1, \hdots, L_{\mathrm{G}}^{\mathrm{t}} \}$. Then, the second-order Taylor expansion of $\bar{\mathsf{P}}_{j}(\mathbf{t}_{\mathrm{B},m})$ w.r.t. $\mathbf{t}_{\mathrm{B},m}$ at the point $\mathbf{t}_{\mathrm{B},m}^{(n)}$ can be expressed as
\begin{align}
\bar{\mathsf{P}}_{j}(\mathbf{t}_{\mathrm{B},m}) &\approx \nabla_{\mathbf{t}_{\mathrm{B},m}}^{\mathsf{T}}(\bar{\mathsf{P}}_{j}(\mathbf{t}_{\mathrm{B},m}))|_{\mathbf{t}_{\mathrm{B},m} = \mathbf{t}_{\mathrm{B},m}^{(n)}}(\mathbf{t}_{\mathrm{B},m} - \mathbf{t}_{\mathrm{B},m}^{(n)}) \notag\\
& + (\mathbf{t}_{\mathrm{B},m} - \mathbf{t}_{\mathrm{B},m}^{(n)})^{\mathsf{T}}\frac{\nabla_{\mathbf{t}_{\mathrm{B},m}}^{2}(\bar{\mathsf{P}}_{j}(\mathbf{t}_{\mathrm{B},m}))|_{\mathbf{t}_{\mathrm{B},m} = \mathbf{t}_{\mathrm{B},m}^{(n)}}}{2} \notag\\
& \times (\mathbf{t}_{\mathrm{B},m} - \mathbf{t}_{\mathrm{B},m}^{(n)}) + \bar{\mathsf{P}}_{j}(\mathbf{t}_{\mathrm{B},m}^{(n)}), \; j \in \mathcal{K}_{\mathrm{E}},
\end{align}
where the gradient $\nabla_{\mathbf{t}_{\mathrm{B},m}}(\bar{\mathsf{P}}_{j}(\mathbf{t}_{\mathrm{B},m}))$ and Hessian matrix $\nabla_{\mathbf{t}_{\mathrm{B},m}}^{2}(\bar{\mathsf{P}}_{j}(\mathbf{t}_{\mathrm{B},m}))$ are respectively given as follows
\begin{align}
&\nabla_{\mathbf{t}_{\mathrm{B},m}}(\bar{\mathsf{P}}_{j}(\mathbf{t}_{\mathrm{B},m})) = \bigg[\frac{\partial\bar{\mathsf{P}}_{j}(\mathbf{t}_{\mathrm{B},m})}{\partial[\mathbf{t}_{\mathrm{B},m}]_{1}}, \frac{\partial\bar{\mathsf{P}}_{j}(\mathbf{t}_{\mathrm{B},m})}{\partial[\mathbf{t}_{\mathrm{B},m}]_{2}}\bigg]^{\mathsf{T}}, \notag\\
&\frac{\partial\bar{\mathsf{P}}_{j}(\mathbf{t}_{\mathrm{B},m})}{\partial[\mathbf{t}_{\mathrm{B},m}]_{1}} = -2\sum_{l=1}^{L_{\mathrm{G}}^{\mathrm{t}}}|[\mathbf{b}_{1,j,m}]_{l}|\frac{2\pi}{\lambda_{\mathrm{c}}}[\bm{\rho}_{\mathrm{G},l}^{\mathrm{t}}]_{1}\mathsf{sin}(\kappa_{j,m,l}^{\mathrm{P}}), \notag\\
&\frac{\partial\bar{\mathsf{P}}_{j}(\mathbf{t}_{\mathrm{B},m})}{\partial[\mathbf{t}_{\mathrm{B},m}]_{2}} = -2\sum_{l=1}^{L_{\mathrm{G}}^{\mathrm{t}}}|[\mathbf{b}_{1,j,m}]_{l}|\frac{2\pi}{\lambda_{\mathrm{c}}}[\bm{\rho}_{\mathrm{G},l}^{\mathrm{t}}]_{2}\mathsf{sin}(\kappa_{j,m,l}^{\mathrm{P}}), \\
&\nabla_{\mathbf{t}_{\mathrm{B},m}}^{2}(\bar{\mathsf{P}}_{j}(\mathbf{t}_{\mathrm{B},m})) = \begin{bmatrix}
\frac{\partial^{2}\bar{\mathsf{P}}_{j}(\mathbf{t}_{\mathrm{B},m})}{\partial[\mathbf{t}_{\mathrm{B},m}]_{1}^{2}} & \frac{\partial^{2}\bar{\mathsf{P}}_{j}(\mathbf{t}_{\mathrm{B},m})}{\partial[\mathbf{t}_{\mathrm{B},m}]_{1}\partial[\mathbf{t}_{\mathrm{B},m}]_{2}} \\
\frac{\partial^{2}\bar{\mathsf{P}}_{j}(\mathbf{t}_{\mathrm{B},m})}{\partial[\mathbf{t}_{\mathrm{B},m}]_{2}\partial[\mathbf{t}_{\mathrm{B},m}]_{1}} & \frac{\partial^{2}\bar{\mathsf{P}}_{j}(\mathbf{t}_{\mathrm{B},m})}{\partial[\mathbf{t}_{\mathrm{B},m}]_{2}^{2}} \\
\end{bmatrix}, \notag\\
&\frac{\partial^{2}\bar{\mathsf{P}}_{j}(\mathbf{t}_{\mathrm{B},m})}{\partial[\mathbf{t}_{\mathrm{B},m}]_{1}^{2}} = -2\sum_{l=1}^{L_{\mathrm{G}}^{\mathrm{t}}}|[\mathbf{b}_{1,j,m}]_{l}|\frac{4\pi^{2}}{\lambda_{\mathrm{c}}^{2}}[\bm{\rho}_{\mathrm{G},l}^{\mathrm{t}}]_{1}^{2}\mathsf{cos}(\kappa_{j,m,l}^{\mathrm{P}}), \notag\\
&\frac{\partial^{2}\bar{\mathsf{P}}_{j}(\mathbf{t}_{\mathrm{B},m})}{\partial[\mathbf{t}_{\mathrm{B},m}]_{1}\partial[\mathbf{t}_{\mathrm{B},m}]_{2}} = -2\sum_{l=1}^{L_{\mathrm{G}}^{\mathrm{t}}}|[\mathbf{b}_{1,j,m}]_{l}|\frac{4\pi^{2}}{\lambda_{\mathrm{c}}^{2}}[\bm{\rho}_{\mathrm{G},l}^{\mathrm{t}}]_{1} \notag\\
&\qquad\qquad\qquad\quad\;\; \times [\bm{\rho}_{\mathrm{G},l}^{\mathrm{t}}]_{2}\mathsf{cos}(\kappa_{j,m,l}^{\mathrm{P}}), \notag\\
&\frac{\partial^{2}\bar{\mathsf{P}}_{j}(\mathbf{t}_{\mathrm{B},m})}{\partial[\mathbf{t}_{\mathrm{B},m}]_{2}\partial[\mathbf{t}_{\mathrm{B},m}]_{1}} = \frac{\partial^{2}\bar{\mathsf{P}}_{j}(\mathbf{t}_{\mathrm{B},m})}{\partial[\mathbf{t}_{\mathrm{B},m}]_{1}\partial[\mathbf{t}_{\mathrm{B},m}]_{2}}, \notag\\
&\frac{\partial^{2}\bar{\mathsf{P}}_{j}(\mathbf{t}_{\mathrm{B},m})}{\partial[\mathbf{t}_{\mathrm{B},m}]_{2}^{2}} = -2\sum_{l=1}^{L_{\mathrm{G}}^{\mathrm{t}}}|[\mathbf{b}_{1,j,m}]_{l}|\frac{4\pi^{2}}{\lambda_{\mathrm{c}}^{2}}[\bm{\rho}_{\mathrm{G},l}^{\mathrm{t}}]_{2}^{2}\mathsf{cos}(\kappa_{j,m,l}^{\mathrm{P}}).
\end{align}
Furthermore, it can be observed that $\nabla_{\mathbf{t}_{\mathrm{B},m}}^{2}(\bar{\mathsf{P}}_{j}(\mathbf{t}_{\mathrm{B},m}))$ has bounded curvature, which indicates that we can acquire a negative semidefinite matrix $\mathbf{M}_{j,m}^{\mathrm{P}}$ such that $\mathbf{M}_{j,m}^{\mathrm{P}} \preceq \nabla_{\mathbf{t}_{\mathrm{B},m}}^{2}(\bar{\mathsf{P}}_{j}(\mathbf{t}_{\mathrm{B},m}))$. Detailed derivation of $\mathbf{M}_{j,m}^{\mathrm{P}}$ is presented in Appendix B. Finally, after substituting $\mathbf{M}_{j,m}^{\mathrm{P}}$ into (53), a concave lower bound of $\mathsf{P}_{j}$ w.r.t. $\mathbf{t}_{\mathrm{B},m}$ can be obtained, which reads
\begin{align}
\mathsf{P}_{j}(\mathbf{t}_{\mathrm{B},m}|\mathbf{t}_{\mathrm{B},m}^{(n)}) &= \nabla_{\mathbf{t}_{\mathrm{B},m}}^{\mathsf{T}}(\bar{\mathsf{P}}_{j}(\mathbf{t}_{\mathrm{B},m}))|_{\mathbf{t}_{\mathrm{B},m} = \mathbf{t}_{\mathrm{B},m}^{(n)}}(\mathbf{t}_{\mathrm{B},m} - \mathbf{t}_{\mathrm{B},m}^{(n)}) \notag\\
& + (\mathbf{t}_{\mathrm{B},m} - \mathbf{t}_{\mathrm{B},m}^{(n)})^{\mathsf{T}}\frac{\mathbf{M}_{j,m}^{\mathrm{P}}}{2}(\mathbf{t}_{\mathrm{B},m} - \mathbf{t}_{\mathrm{B},m}^{(n)}) \notag\\
& + (\mathbf{g}_{j}^{(n)})^{\mathsf{H}}\mathbf{F}\mathbf{g}_{j}^{(n)}, \; j \in \mathcal{K}_{\mathrm{E}}.
\end{align}

Lastly, we take (37c) into consideration, which is much easier to handle since its left hand side is convex w.r.t. $\mathbf{t}_{\mathrm{B},m}$. Therefore, by adopting its first-order Taylor expansion w.r.t. $\mathbf{t}_{\mathrm{B},m}$ at the point $\mathbf{t}_{\mathrm{B},m}^{(n)}$, which is given by
\begin{align}
\| \mathbf{t}_{\mathrm{B},m} - \mathbf{t}_{\mathrm{B},s} \|_{2} &\geq \frac{(\mathbf{t}_{\mathrm{B},m}^{(n)} - \mathbf{t}_{\mathrm{B},s})^{\mathsf{T}}}{\| \mathbf{t}_{\mathrm{B},m}^{(n)} - \mathbf{t}_{\mathrm{B},s} \|_{2}}(\mathbf{t}_{\mathrm{B},m} - \mathbf{t}_{\mathrm{B},s}) \notag\\
& \triangleq \mathsf{D}_{s}(\mathbf{t}_{\mathrm{B},m}|\mathbf{t}_{\mathrm{B},m}^{(n)}), \; s \in \mathcal{M}, \; m \neq s,
\end{align}
and surrogating $\mathsf{D}_{s}(\mathbf{t}_{\mathrm{B},m}|\mathbf{t}_{\mathrm{B},m}^{(n)})$ into (37c), (37c) can thus be convexified.

\begin{algorithm}[!t]
\caption{Solution to $(\mathcal{P}2)$}
\begin{algorithmic}[1]
\STATE Initialize feasible $\tilde{\mathbf{f}}^{(0)}$, $\tilde{\mathbf{t}}_{\mathrm{B}}^{(0)}$, $\bm{\theta}^{(0)}$, and $n = 0$;
\REPEAT
\STATE update $\mathbf{v}^{(n + 1)}$ by (21);
\STATE update $\mathbf{w}^{(n + 1)}$ by (23);
\STATE update $\tilde{\mathbf{f}}^{(n + 1)}$ by solving $(\mathcal{P}3)$;
\STATE update $\bm{\theta}^{(n + 1)}$ by Alg. 1;
\FOR{$m = 1$ to $M$}
\STATE update $\mathbf{t}_{\mathrm{B},m}^{(n + 1)}$ by solving $(\mathcal{P}5)$;
\ENDFOR
\STATE set $n := n + 1$;
\UNTIL{convergence}
\end{algorithmic}
\end{algorithm}

According to the above developed convex surrogate (49), and concave surrogates (56) and (57), a tight approximated convex optimization problem w.r.t. $\mathbf{t}_{\mathrm{B},m}$ can be formulated as
\begin{align}
(\mathcal{P}5): \min_{\mathbf{t}_{\mathrm{B},m}} \; &\tilde{\mathsf{R}}(\mathbf{t}_{\mathrm{B},m}|\mathbf{t}_{\mathrm{B},m}^{(n)}) \\
\mathrm{s.t.} \; &\mathsf{P}_{j}(\mathbf{t}_{\mathrm{B},m}|\mathbf{t}_{\mathrm{B},m}^{(n)}) \geq P_{\mathrm{E},j}, \; j \in \mathcal{K}_{\mathrm{E}}, \tag{58a}\\
&\mathbf{t}_{\mathrm{B},m} \in \mathcal{C}_{\mathrm{B}}, \tag{58b}\\
&\mathsf{D}_{s}(\mathbf{t}_{\mathrm{B},m}|\mathbf{t}_{\mathrm{B},m}^{(n)}) \geq D_{\mathrm{B}}, \; s \in \mathcal{M}, \; m \neq s, \tag{58c}
\end{align}
which can be solved by numerical solvers such as CVX.

The proposed algorithm for solving $(\mathcal{P}2)$ (equivalent to solving $(\mathcal{P}1)$) is summarized in Algorithm 2.

\section{Feasibility Characterization}

It is worth noting that, in the last section, we propose Alg. 1 under the assumption that the feasible domain of $(\mathcal{P}1)$ is non-empty. However, how to characterize its feasibility is still absent. Hence, in this section, we present a method to tackle this problem.

\subsection{Problem Formulation}

The feasibility characterization problem of $(\mathcal{P}1)$ is given by
\begin{align}
\mathsf{Find} \; &(\tilde{\mathbf{f}}, \tilde{\mathbf{t}}_{\mathrm{B}}, \bm{\theta}) \\
\mathrm{s.t.} \; &\mathsf{P}_{j} \geq P_{\mathrm{E},j}, \; j \in \mathcal{K}_{\mathrm{E}}, \tag{59a}\\
&\sum_{i=1}^{K_{\mathrm{I}}}\| \mathbf{f}_{\mathrm{I},i} \|_{2}^{2} + \sum_{j=1}^{K_{\mathrm{E}}}\| \mathbf{f}_{\mathrm{E},j} \|_{2}^{2} \leq P_{\mathrm{B}}, \tag{59b}\\
&0 \leq \theta_{n} \leq 2\pi, \; n \in \mathcal{N}, \tag{59c}\\
&\mathbf{t}_{\mathrm{B},m} \in \mathcal{C}_{\mathrm{B}}, \; m \in \mathcal{M}, \tag{59d}\\
&\| \mathbf{t}_{\mathrm{B},m} - \mathbf{t}_{\mathrm{B},s} \|_{2} \geq D_{\mathrm{B}}, \; m, s \in \mathcal{M}, \; m \neq s, \tag{59e}
\end{align}
which is troublesome due to the fact that its objective is missing. Therefore, we consider another relevant problem, which can be expressed as
\begin{align}
(\mathcal{P}6): &\min_{\tilde{\mathbf{f}}, \tilde{\mathbf{t}}_{\mathrm{B}}, \bm{\theta}, \beta} \; \beta \\
\mathrm{s.t.} \; &P_{\mathrm{E},j} - \mathsf{P}_{j} \leq \beta, \; j \in \mathcal{K}_{\mathrm{E}}, \tag{60a}\\
&\sum_{i=1}^{K_{\mathrm{I}}}\| \mathbf{f}_{\mathrm{I},i} \|_{2}^{2} + \sum_{j=1}^{K_{\mathrm{E}}}\| \mathbf{f}_{\mathrm{E},j} \|_{2}^{2} \leq P_{\mathrm{B}}, \tag{60b}\\
&0 \leq \theta_{n} \leq 2\pi, \; n \in \mathcal{N}, \tag{60c}\\
&\mathbf{t}_{\mathrm{B},m} \in \mathcal{C}_{\mathrm{B}}, \; m \in \mathcal{M}, \tag{60d}\\
&\| \mathbf{t}_{\mathrm{B},m} - \mathbf{t}_{\mathrm{B},s} \|_{2} \geq D_{\mathrm{B}}, \; m, s \in \mathcal{M}, \; m \neq s. \tag{60e}
\end{align}

Denote $(\tilde{\mathbf{f}}^{\star}, \tilde{\mathbf{t}}_{\mathrm{B}}^{\star}, \bm{\theta}^{\star}, \beta^{\star})$ as the optimal solution to $(\mathcal{P}6)$. If $\beta^{\star} \leq 0$, then $(\mathcal{P}1)$ is feasible and the associated $(\tilde{\mathbf{f}}^{\star}, \tilde{\mathbf{t}}_{\mathrm{B}}^{\star}, \bm{\theta}^{\star})$ is actually a feasible point of $(\mathcal{P}1)$. Otherwise, $(\mathcal{P}1)$ is infeasible. As a result, we need to develop an algorithm for solving $(\mathcal{P}6)$, which is detailed in the following.

\subsection{Solution to $(\mathcal{P}6)$}

We again utilize BCD method to tackle $(\mathcal{P}6)$, where in each block, we update $\beta$ together with one of the variables $\tilde{\mathbf{f}}$, $\tilde{\mathbf{t}}_{\mathrm{B}}$, and $\bm{\theta}$ when other variables are given.

\subsubsection{Updating $(\tilde{\mathbf{f}}, \beta)$} With other variables being fixed, the subproblem w.r.t. $(\tilde{\mathbf{f}}, \beta)$ is given as follows
\begin{align}
\min_{\tilde{\mathbf{f}}, \beta} \; &\beta \\
\mathrm{s.t.} \; &P_{\mathrm{E},j} - \mathsf{P}_{j} \leq \beta, \; j \in \mathcal{K}_{\mathrm{E}}, \tag{61a}\\
&\sum_{i=1}^{K_{\mathrm{I}}}\| \mathbf{f}_{\mathrm{I},i} \|_{2}^{2} + \sum_{j=1}^{K_{\mathrm{E}}}\| \mathbf{f}_{\mathrm{E},j} \|_{2}^{2} \leq P_{\mathrm{B}}, \tag{61b}
\end{align}
which is difficult to solve since (61a) is nonconvex. To overcome this difficulty, we substitute $\mathsf{P}_{j}$'s concave lower bound $\mathsf{P}_{j}(\tilde{\mathbf{f}}|\tilde{\mathbf{f}}^{(n)})$ derived in (25) into (61a) to obtain a convex optimization problem, written as
\begin{align}
(\mathcal{P}7): \min_{\tilde{\mathbf{f}}, \beta} \; &\beta \\
\mathrm{s.t.} \; &P_{\mathrm{E},j} - \mathsf{P}_{j}(\tilde{\mathbf{f}}|\tilde{\mathbf{f}}^{(n)}) \leq \beta, \; j \in \mathcal{K}_{\mathrm{E}}, \tag{62a}\\
&\sum_{i=1}^{K_{\mathrm{I}}}\| \mathbf{f}_{\mathrm{I},i} \|_{2}^{2} + \sum_{j=1}^{K_{\mathrm{E}}}\| \mathbf{f}_{\mathrm{E},j} \|_{2}^{2} \leq P_{\mathrm{B}}, \tag{62b}
\end{align}
which can be numerically solved.

\subsubsection{Updating $(\bm{\theta}, \beta)$}

Next, we take the update of $(\bm{\theta}, \beta)$ into account, whose optimization problem can be casted as
\begin{align}
\min_{\bm{\theta}, \beta} \; &\beta \\
\mathrm{s.t.} \; &P_{\mathrm{E},j} - \bm{\theta}^{\mathsf{H}}\mathbf{Q}_{1,j}\bm{\theta} \leq \beta, \; j \in \mathcal{K}_{\mathrm{E}}, \tag{63a}\\
&0 \leq \theta_{n} \leq 2\pi, \; n \in \mathcal{N}, \tag{63b}
\end{align}
where $\mathbf{Q}_{1,j}, \; j \in \mathcal{K}_{\mathrm{E}}$, has been given in (28). This problem is challenging to handle due to the nonconvex constraint (63a) and the constant modulus form of (63b). We again exploit PDD framework [32] to tackle this problem. Similar to updating $\bm{\theta}$ shown in the last section, via introducing an auxiliary variable $\bm{\phi}$ and penalizing the equality constraint, we turn to solve the following augmented Lagrangian problem
\begin{align}
\min_{\bm{\theta}, \bm{\phi}, \beta} \; &\beta + \frac{1}{2\rho}\| \bm{\theta} - \bm{\phi} + \rho\bm{\lambda} \|_{2}^{2} \\
\mathrm{s.t.} \; &P_{\mathrm{E},j} - \bm{\theta}^{\mathsf{H}}\mathbf{Q}_{1,j}\bm{\theta} \leq \beta, \; j \in \mathcal{K}_{\mathrm{E}}, \tag{64a}\\
&|[\bm{\theta}]_{n}| \leq 1, \; n \in \mathcal{N}, \tag{64b}\\
&|[\bm{\phi}]_{n}| = 1, \; n \in \mathcal{N}, \tag{64c}
\end{align}
where $\rho$ and $\bm{\lambda}$ are similarly defined. To avoid repetition, we directly present the two-layer procedure based on PDD method.

\begin{algorithm}[!t]
\caption{PDD-Based Solution to $(\bm{\theta}, \beta)$}
\begin{algorithmic}[1]
\STATE Initialize $\bm{\theta}^{(0)}$, $\bm{\phi}^{(0)}$, $\beta^{(0)}$, $\bm{\lambda}^{(0)}$, $\rho^{(0)}$, and $n = 0$;
\REPEAT
\STATE set $\bm{\theta}^{(n, 0)} := \bm{\theta}^{(n)}$, $\bm{\phi}^{(n, 0)} := \bm{\phi}^{(n)}$, $\beta^{(n, 0)} := \beta^{(n)}$, \\ and $i = 0$;
\REPEAT
\STATE update $(\bm{\theta}^{(n, i + 1)}, \beta^{(n, i + 1)})$ by solving $(\mathcal{P}8)$;
\STATE update $\bm{\phi}^{(n, i + 1)}$ by (66);
\STATE set $i := i + 1$;
\UNTIL{convergence}
\STATE set $\bm{\theta}^{(n + 1)} := \bm{\theta}^{(n, \infty)}$, $\bm{\phi}^{(n + 1)} := \bm{\phi}^{(n, \infty)}$, and $\beta^{(n + 1)} := \beta^{(n, \infty)}$;
\IF{$\| \bm{\theta}^{(n + 1)} - \bm{\phi}^{(n + 1)} \|_{\infty} \leq \delta$}
\STATE $\bm{\lambda}^{(n + 1)} := \bm{\lambda}^{(n)} + \frac{1}{\rho^{(n)}}(\bm{\theta}^{(n + 1)} - \bm{\phi}^{(n + 1)})$, $\rho^{(n + 1)} := \rho^{(n)}$;
\ELSE
\STATE $\bm{\lambda}^{(n + 1)} := \bm{\lambda}^{(n)}$, $\rho^{(n + 1)} := c_{\rho} \cdot \rho^{(n)}$;
\ENDIF
\STATE set $n := n + 1$;
\UNTIL{$\| \bm{\theta}^{(n)} - \bm{\phi}^{(n)} \|_{2}$ is sufficiently small}
\end{algorithmic}
\end{algorithm}

For the inner layer, when $\bm{\phi}$ is fixed, after surrogating $\mathsf{P}_{j}(\bm{\theta}|\bm{\theta}^{(n)})$ shown in (32) into (64a), the subproblem w.r.t. $(\bm{\theta}, \beta)$ reads
\begin{align}
(\mathcal{P}8): \min_{\bm{\theta}, \beta} \; &\beta + \frac{1}{2\rho}\| \bm{\theta} - \bm{\phi} + \rho\bm{\lambda} \|_{2}^{2} \\
\mathrm{s.t.} \; &P_{\mathrm{E},j} - \mathsf{P}_{j}(\bm{\theta}|\bm{\theta}^{(n)}) \leq \beta, \; j \in \mathcal{K}_{\mathrm{E}}, \tag{65a}\\
&|[\bm{\theta}]_{n}| \leq 1, \; n \in \mathcal{N}, \tag{65b}
\end{align}
which is convex and can be solved by CVX.

When $(\bm{\theta}, \beta)$ is fixed, the optimal solution to $\bm{\phi}$ can be expressed in a closed form as
\begin{align}
\bm{\phi} = e^{\jmath\angle(\bm{\theta} + \rho\bm{\lambda})}.
\end{align}

For the outer layer, the Lagrangian multiplier $\bm{\lambda}$ needs to be updated provided that $\bm{\theta}$ sufficiently approaches $\bm{\phi}$. Otherwise, we decrease the value of penalty coefficient $\rho$.

The PDD-based solution to the update of $(\bm{\theta}, \beta)$ is summarized in Algorithm 3.

\subsubsection{Updating $(\tilde{\mathbf{t}}_{\mathrm{B}}, \beta)$} Finally, we proceed to update the positions of MAs and $\beta$. We still optimize one MA's position at a time, which implies to tackle the following optimization
\begin{align}
\min_{\mathbf{t}_{\mathrm{B},m}, \beta} \; &\beta \\
\mathrm{s.t.} \; &P_{\mathrm{E},j} - \mathsf{P}_{j} \leq \beta, \; j \in \mathcal{K}_{\mathrm{E}}, \tag{67a}\\
&\mathbf{t}_{\mathrm{B},m} \in \mathcal{C}_{\mathrm{B}}, \tag{67b}\\
&\| \mathbf{t}_{\mathrm{B},m} - \mathbf{t}_{\mathrm{B},s} \|_{2} \geq D_{\mathrm{B}}, \; s \in \mathcal{M}, \; m \neq s, \tag{67c}
\end{align}
which is nonconvex resulting from (67a) and (67c). By recalling the constructed concave lower bounds (56) and (57), we acquire a tight approximated optimization problem, given by
\begin{align}
(\mathcal{P}9): \min_{\mathbf{t}_{\mathrm{B},m}, \beta} \; &\beta \\
\mathrm{s.t.} \; &P_{\mathrm{E},j} - \mathsf{P}_{j}(\mathbf{t}_{\mathrm{B},m}|\mathbf{t}_{\mathrm{B},m}^{(n)}) \leq \beta, \; j \in \mathcal{K}_{\mathrm{E}}, \tag{68a}\\
&\mathbf{t}_{\mathrm{B},m} \in \mathcal{C}_{\mathrm{B}}, \tag{68b}\\
&\mathsf{D}_{s}(\mathbf{t}_{\mathrm{B},m}|\mathbf{t}_{\mathrm{B},m}^{(n)}) \geq D_{\mathrm{B}}, \; s \in \mathcal{M}, \; m \neq s. \tag{68c}
\end{align}
Problem $(\mathcal{P}9)$ is convex which can be numerically solved.

The proposed algorithm for checking the feasibility of $(\mathcal{P}1)$ (i.e., solving $(\mathcal{P}6)$) is stated in Algorithm 4.

\begin{algorithm}[!t]
\caption{Feasibility Characterization for $(\mathcal{P}1)$}
\begin{algorithmic}[1]
\STATE Initialize $\tilde{\mathbf{f}}^{(0)}$, $\tilde{\mathbf{t}}_{\mathrm{B}}^{(0)}$, $\bm{\theta}^{(0)}$, and $n = 0$;
\REPEAT
\STATE update $(\tilde{\mathbf{f}}^{(n + 1)}, \beta^{(n + \frac{1}{M + 2})})$ by solving $(\mathcal{P}7)$;
\STATE update $(\bm{\theta}^{(n + 1)}, \beta^{(n + \frac{2}{M + 2})})$ by Alg. 3;
\FOR{$m = 1$ to $M$}
\STATE update $(\mathbf{t}_{\mathrm{B},m}^{(n + 1)}, \beta^{(n + \frac{m + 2}{M + 2})})$ by solving $(\mathcal{P}9)$;
\ENDFOR
\STATE set $n := n + 1$;
\UNTIL{convergence}
\IF{$\beta^{\star} \leq 0$}
\STATE $(\mathcal{P}1)$ is \textbf{feasible};
\ELSE
\STATE $(\mathcal{P}1)$ is \textbf{infeasible};
\ENDIF
\end{algorithmic}
\end{algorithm}

\section{Simulation Results}

In this section, we provide numerical results to demonstrate the effectiveness of our proposed algorithms and the benefit of applying an IRS and MAs to SWIPT system. Unless otherwise specified, $P_{\mathrm{B}} = 40$ dBm, $M = 4$, $N = 16$, $K_{\mathrm{I}} = 3$, $K_{\mathrm{E}} = 3$, $\lambda_{\mathrm{c}} = 0.125$ m, $A = 2.5\lambda_{\mathrm{c}}$, $D_{\mathrm{B}} = 0.5\lambda_{\mathrm{c}}$, $\alpha_{i} = 1$, $i \in \mathcal{K}_{\mathrm{I}}$, $\sigma_{\mathrm{I},i}^{2} = -90$ dBm, $i \in \mathcal{K}_{\mathrm{I}}$, $\rho = 0.5$, $c_{\rho} = 0.75$.

In the simulations, the number of transmit paths and that of receive paths for all channels are set as $L = 5$. The path response matrices $\bm{\Sigma}_{\mathrm{G}}$, $\bm{\Sigma}_{\mathrm{h},i}, \; i \in \mathcal{K}_{\mathrm{I}}$, and $\bm{\Sigma}_{\mathrm{g},j}, \; j \in \mathcal{K}_{\mathrm{E}}$, are all diagonal matrices [16], with $[\bm{\Sigma}_{\mathrm{G}}]_{l,l} \sim \mathcal{CN}(0, \frac{C_{0}^{2}d_{\mathrm{G}}^{-\alpha_{\mathrm{G}}}}{L}), \; l \in \{ 1, \hdots, L \}$, $[\bm{\Sigma}_{\mathrm{h},i}]_{l,l} \sim \mathcal{CN}(0, \frac{C_{0}^{2}d_{\mathrm{h},i}^{-\alpha_{\mathrm{h},i}}}{L}), \; i \in \mathcal{K}_{\mathrm{I}}, \; l \in \{ 1, \hdots, L \}$, and $[\bm{\Sigma}_{\mathrm{g},j}]_{l,l} \sim \mathcal{CN}(0, \frac{C_{0}^{2}d_{\mathrm{g},j}^{-\alpha_{\mathrm{g},j}}}{L}), \; j \in \mathcal{K}_{\mathrm{E}}, \; l \in \{ 1, \hdots, L \}$, where $C_{0} = \frac{\lambda_{\mathrm{c}}}{4\pi}$ represents the path loss with reference distance of $1$ m, $d_{\mathrm{G}}$, $d_{\mathrm{h},i}, \; i \in \mathcal{K}_{\mathrm{I}}$, and $d_{\mathrm{g},j}, \; j \in \mathcal{K}_{\mathrm{E}}$, denote the distances between the BS and IRS, between the IRS and the $i$-th IDR, and between the IRS and the $j$-th EHR, respectively, $\alpha_{\mathrm{G}}$, $\alpha_{\mathrm{h},i}, \; i \in \mathcal{K}_{\mathrm{I}}$, and $\alpha_{\mathrm{g},j}, \; j \in \mathcal{K}_{\mathrm{E}}$, are the path loss exponents of $\mathbf{G}$, $\mathbf{h}_{i}$, and $\mathbf{g}_{j}$, respectively. Unless otherwise specified, $d_{\mathrm{G}} = 4$ m, $d_{\mathrm{h},i} \in [20, 25]$ m, $i \in \mathcal{K}_{\mathrm{I}}$, $d_{\mathrm{g},j} \in [4, 4.5]$ m, $j \in \mathcal{K}_{\mathrm{E}}$, and all the path loss exponents are set as $2.2$ [25].

Fig. 2 illustrates the convergence behavior of our proposed algorithm for solving $(\mathcal{P}2)$. As suggested by this figure, by utilizing Alg. 2, the objective value of $(\mathcal{P}2)$ monotonically increases and converges within tens of iterations. Besides, the objective value of $(\mathcal{P}2)$ improves with the growth of BS power budget $P_{\mathrm{B}}$.

\begin{figure}[!t]
\centering
\includegraphics[scale=0.53]{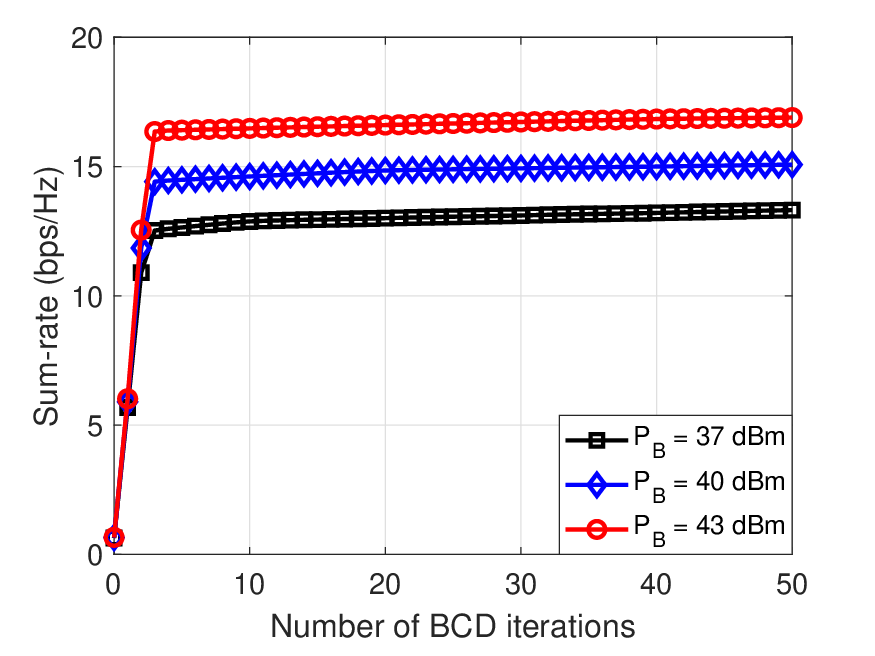}
\caption*{Fig. 2. Convergence behavior of Alg. 2.}
\end{figure}

\begin{figure}[!t]
\centering
\includegraphics[scale=0.53]{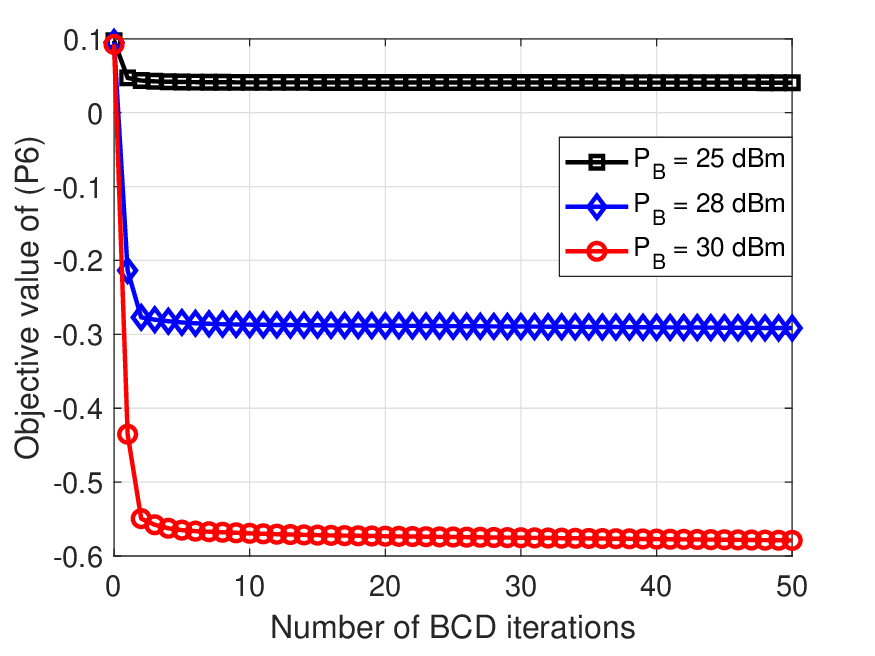}
\caption*{Fig. 3. Convergence behavior of Alg. 4.}
\end{figure}

Fig. 3 plots the convergence behavior of our proposed feasibility characterization method. It can be observed from this figure that, Alg. 4 guarantees monotonically decreasing objective values of $(\mathcal{P}6)$ and exhibits fast convergence. Note that when $P_{\mathrm{B}}$ reduces, the objective value of $(\mathcal{P}6)$ increases, which indicates that the feasible domain of $(\mathcal{P}1)$ shrinks. If $P_{\mathrm{B}}$ is sufficiently small, $(\mathcal{P}1)$ can even become infeasible, as shown by the black curve in this figure.

In Fig. 4, we investigate the impact of BS maximal transmit power $P_{\mathrm{B}}$ on the sum-rate of IDRs. Besides our proposed solution labelled as “\textbf{MA-OPS}”, i.e., the BS is equipped with MAs and the phase shifts of IRS are optimized, we also consider other schemes for comparison: i) “\textbf{FPA-OPS}”: the IRS phase shifts are optimized while the BS antennas cannot move; ii) “\textbf{MA-RPS}”: MAs are deployed at the BS while the phase shifts of IRS are randomly configured; iii) “\textbf{FPA-RPS}”: the BS is equipped with FPAs and random IRS phase shifts are adopted. As reflected by this figure, the sum-rate of IDRs can be boosted via increasing $P_{\mathrm{B}}$. Furthermore, our proposed solution is superior over all other considered schemes, which verifies the necessity of adjusting the positions of BS antennas and optimizing the IRS configuration. Interestingly, under our simulation settings, compared to the “FPA-RPS” scheme, the performance improvement yielded by “FPA-OPS” scheme is larger than “MA-RPS” counterpart. Meanwhile, compared to our proposed “MA-OPS” scheme, “MA-RPS” scheme exhibits a larger performance gap than “FPA-OPS” scheme. Both phenomena indicate that the optimized IRS configuration contributes more to our considered scenario than MAs.

\begin{figure}[!t]
\centering
\includegraphics[scale=0.53]{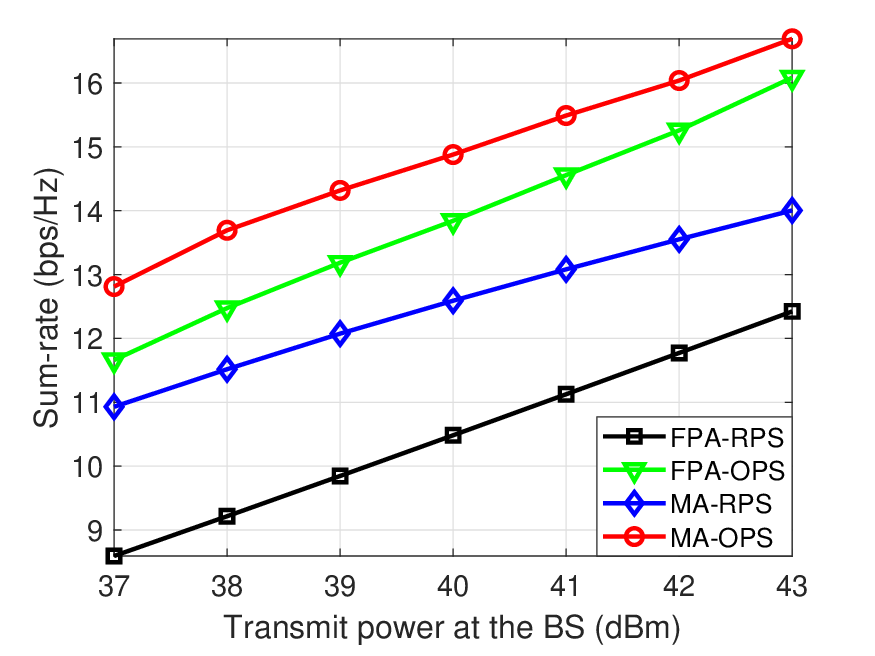}
\caption*{Fig. 4. Impact of BS power budget $P_{\mathrm{B}}$ on sum-rate.}
\end{figure}

\begin{figure}[!t]
\centering
\includegraphics[scale=0.53]{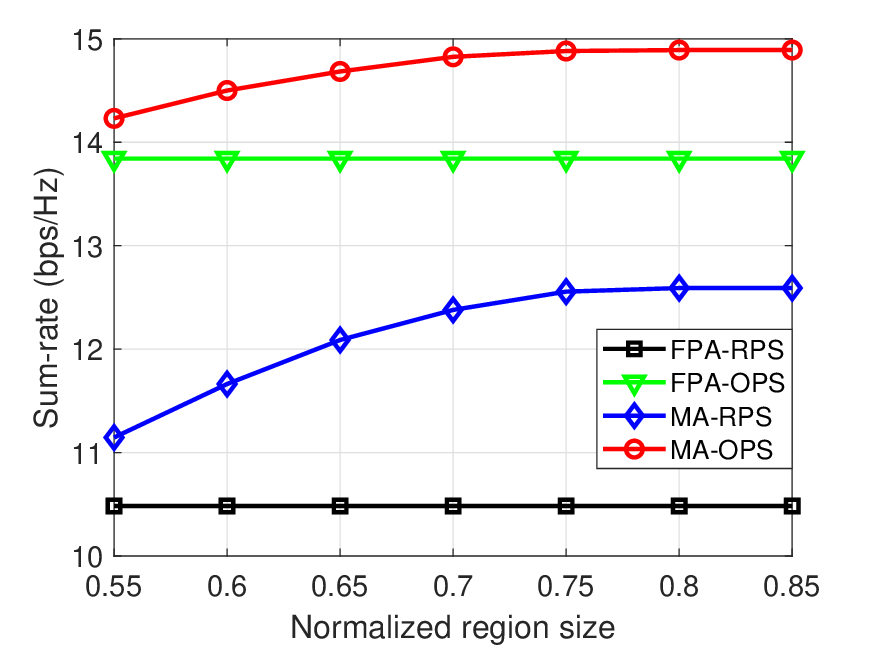}
\caption*{Fig. 5. Impact of normalized BS array size on sum-rate.}
\end{figure}

Fig. 5 illustrates the impact of normalized BS array size (defined as $A/\lambda_{\mathrm{c}}$) on information transmission. It can be seen from this figure that, for “MA-OPS” and “MA-RPS” schemes, the system performance improves with the BS array size becoming larger since the feasible domain of $(\mathcal{P}1)$ expands. However, when the BS array size is adequately large, enlarging array size only brings marginal performance improvement due to the fact that the optimal MAs' positions vary negligibly little in this case. Besides, our proposed solution achieves the best performance among all considered schemes. In addition, our considered system benefits more from optimizing IRS phase shifts than antenna position optimization.

\begin{figure}[!t]
\centering
\includegraphics[scale=0.53]{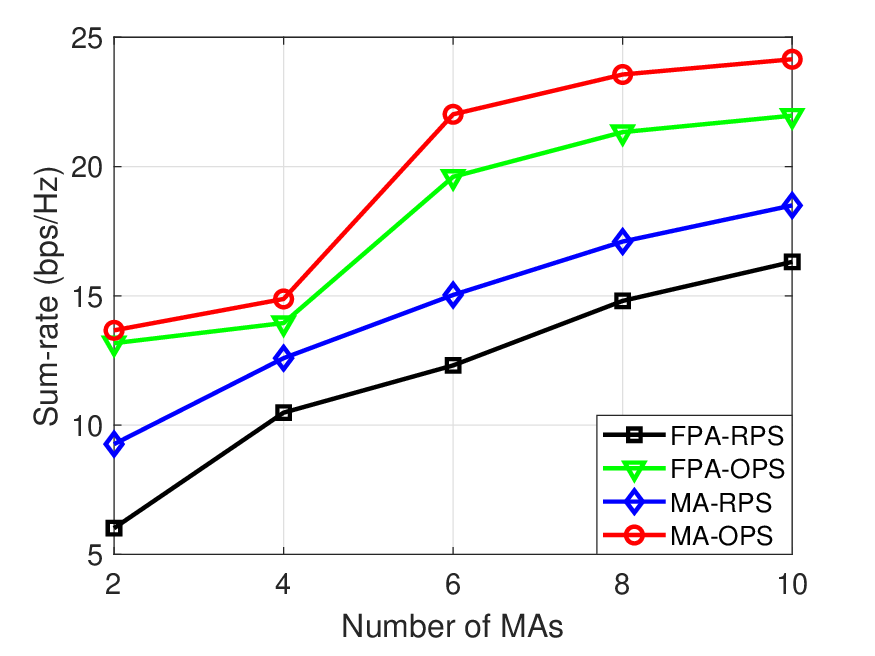}
\caption*{Fig. 6. Sum-rate versus the number of BS antennas $M$.}
\end{figure}

\begin{figure}[!t]
\centering
\includegraphics[scale=0.53]{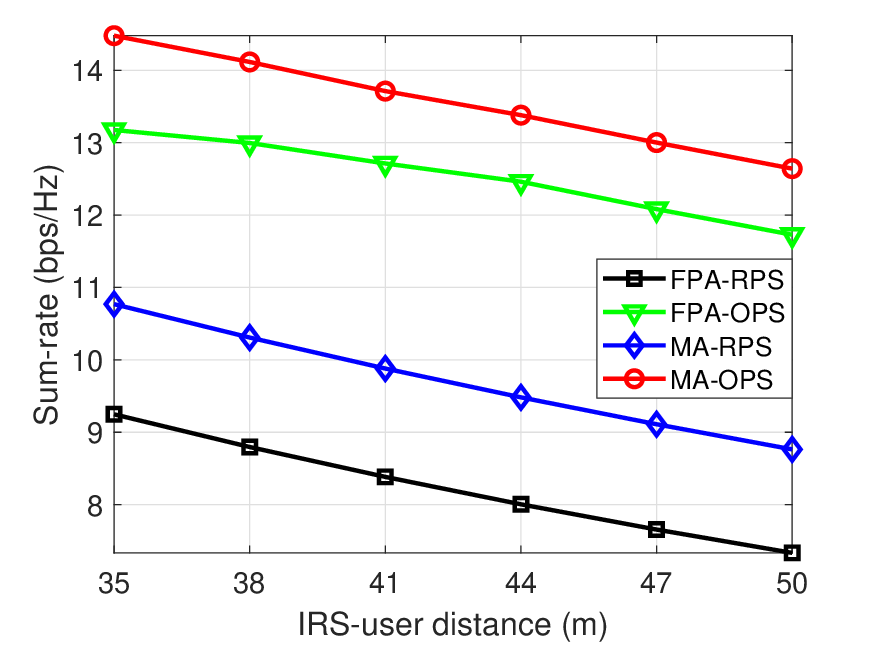}
\caption*{Fig. 7. Impact of distances between the IRS and IDRs on sum-rate.}
\end{figure}

Fig. 6 examines the sum-rate performance influenced by the number of MAs at the BS. As shown by this figure, the larger the number of MAs is, the better the system performance is, thanks to the higher beamforming gain introduced by more MAs. Moreover, our proposed scheme outperforms all other counterparts that are taken into account in this experiment. Besides, the optimized IRS configuration yields more sum-rate increasement than adjusting the positions of MAs in our considered scenario.

Fig. 7 plots the impact of distances between the IRS and IDRs on the sum-rate of data transmission. Here $d_{\mathrm{h},i} \in [\tilde{d}_{\mathrm{h}}, \tilde{d}_{\mathrm{h}} + 5]$ m, $i \in \mathcal{K}_{\mathrm{I}}$, where $\tilde{d}_{\mathrm{h}}$ is the value of the abscissa in Fig. 7. It can be observed from this figure that, the sum-rate performance degrades with the IDRs farther away from the IRS, whose reason lies in much severer path loss between the IRS and each IDR that further undermines the channel gain of cascaded BS-IRS-IDR link. Besides, via appropriately configuring the phase shifts of IRS and the positions of MAs, the system performance can be significantly improved. Again, we draw a conclusion that the optimized IRS configuration has an advantage over proper MA adjustment in enhancing the information transfer of considered system.

\section{Conclusion}

In this paper, we have considered an SWIPT system with an IRS and MAs being employed to assist this system. We have formulated a sum-rate maximization problem via jointly optimizing BS beamforming, MA configuration, and IRS phase shifts. To tackle this challenging optimization, WMMSE method is first utilized for equivalent transformation. Then, BCD framework is exploited to obtain an effective solution, where MM and PDD methods are leveraged to handle the non-convexity of subproblems. Particularly, we have presented a criterion to characterize the feasibility of aforementioned problem. Simulation results have demonstrated that our proposed algorithms exhibit monotonicity and fast convergence, and the performance of SWIPT system can be remarkably improved by both MA and IRS deployment.

\appendix

\subsection{Derivation of $\mathbf{M}_{m}^{\mathrm{R}}, \; m \in \mathcal{M}$}

Denote $H_{1,m}^{\mathrm{R}} \triangleq \frac{\partial^{2}\bar{\mathsf{R}}(\mathbf{t}_{\mathrm{B},m})}{\partial[\mathbf{t}_{\mathrm{B},m}]_{1}^{2}}$, $H_{2,m}^{\mathrm{R}} \triangleq \frac{\partial^{2}\bar{\mathsf{R}}(\mathbf{t}_{\mathrm{B},m})}{\partial[\mathbf{t}_{\mathrm{B},m}]_{1}\partial[\mathbf{t}_{\mathrm{B},m}]_{2}}$, and $H_{3,m}^{\mathrm{R}} \triangleq \frac{\partial^{2}\bar{\mathsf{R}}(\mathbf{t}_{\mathrm{B},m})}{\partial[\mathbf{t}_{\mathrm{B},m}]_{2}^{2}}$. Then, the Hessian matrix $\nabla_{\mathbf{t}_{\mathrm{B},m}}^{2}(\bar{\mathsf{R}}(\mathbf{t}_{\mathrm{B},m}))$ can be rewritten as
\begin{align}
\nabla_{\mathbf{t}_{\mathrm{B},m}}^{2}(\bar{\mathsf{R}}(\mathbf{t}_{\mathrm{B},m})) = \underbrace{\begin{bmatrix}
H_{1,m}^{\mathrm{R}} & 0 \\
0 & H_{3,m}^{\mathrm{R}} \\
\end{bmatrix}}_{\triangleq \mathbf{H}_{1,m}^{\mathrm{R}}} + \underbrace{\begin{bmatrix}
0 & H_{2,m}^{\mathrm{R}} \\
H_{2,m}^{\mathrm{R}} & 0 \\
\end{bmatrix}}_{\triangleq \mathbf{H}_{2,m}^{\mathrm{R}}}.
\end{align}
Hence, the task of constructing $\mathbf{M}_{m}^{\mathrm{R}}$ reduces to exploring the upper bounds of $\mathbf{H}_{1,m}^{\mathrm{R}}$ and $\mathbf{H}_{2,m}^{\mathrm{R}}$, respectively.

Taking $\mathbf{H}_{1,m}^{\mathrm{R}}$ into account first, its upper bound can be obtained by examining the upper bounds of $H_{1,m}^{\mathrm{R}}$ and $H_{3,m}^{\mathrm{R}}$, respectively. Since $\mathsf{cos}(\cdot) \in [-1, 1]$, we have
\begin{align}
H_{1,m}^{\mathrm{R}} &\leq \sum_{l=1}^{L_{\mathrm{G}}^{\mathrm{t}}}|[\mathbf{b}_{0,m}]_{l}|\frac{4\pi^{2}}{\lambda_{\mathrm{c}}^{2}}[\bm{\rho}_{\mathrm{G},l}^{\mathrm{t}}]_{1}^{2}, \\
H_{3,m}^{\mathrm{R}} &\leq \sum_{l=1}^{L_{\mathrm{G}}^{\mathrm{t}}}|[\mathbf{b}_{0,m}]_{l}|\frac{4\pi^{2}}{\lambda_{\mathrm{c}}^{2}}[\bm{\rho}_{\mathrm{G},l}^{\mathrm{t}}]_{2}^{2}.
\end{align}
Therefore, via substituting (70) and (71) into $\mathbf{H}_{1,m}^{\mathrm{R}}$, an upper bound of $\mathbf{H}_{1,m}^{\mathrm{R}}$ can be achieved.

Next, we proceed to derive an upper bound for $\mathbf{H}_{2,m}^{\mathrm{R}}$. Note that $\mathbf{H}_{2,m}^{\mathrm{R}}$ naturally admits the following upper bound
\begin{align}
\mathbf{H}_{2,m}^{\mathrm{R}} \leq |H_{2,m}^{\mathrm{R}}|\mathbf{I}_{2},
\end{align}
which indicates that an upper bound of $\mathbf{H}_{2,m}^{\mathrm{R}}$ can be yielded by checking the upper bound of $|H_{2,m}^{\mathrm{R}}|$. Due to the fact that $\mathsf{cos}(\cdot) \in [-1, 1]$, the upper bound of $|H_{2,m}^{\mathrm{R}}|$ can be given by
\begin{align}
|H_{2,m}^{\mathrm{R}}| \leq \sum_{l=1}^{L_{\mathrm{G}}^{\mathrm{t}}}|[\mathbf{b}_{0,m}]_{l}|\frac{4\pi^{2}}{\lambda_{\mathrm{c}}^{2}}\big|[\bm{\rho}_{\mathrm{G},l}^{\mathrm{t}}]_{1}[\bm{\rho}_{\mathrm{G},l}^{\mathrm{t}}]_{2}\big|.
\end{align}
Consequently, by replacing $|H_{2,m}^{\mathrm{R}}|$ in (72) by (73), we acquire an upper bound of $\mathbf{H}_{2,m}^{\mathrm{R}}$, and the derivation of $\mathbf{M}_{m}^{\mathrm{R}}$ is thus completed.

\subsection{Derivation of $\mathbf{M}_{j,m}^{\mathrm{P}}, \; j \in \mathcal{K}_{\mathrm{E}}, \; m \in \mathcal{M}$}

We first recast the Hessian matrix $\nabla_{\mathbf{t}_{\mathrm{B},m}}^{2}(\bar{\mathsf{P}}_{j}(\mathbf{t}_{\mathrm{B},m}))$ as
\begin{align}
\nabla_{\mathbf{t}_{\mathrm{B},m}}^{2}(\bar{\mathsf{P}}_{j}(\mathbf{t}_{\mathrm{B},m})) \!\!=\!\! \underbrace{\begin{bmatrix}
H_{1,j,m}^{\mathrm{P}} & \!\!\!\!0 \\
0 & \!\!\!\!H_{3,j,m}^{\mathrm{P}} \\
\end{bmatrix}}_{\triangleq \mathbf{H}_{1,j,m}^{\mathrm{P}}} \!\!+\!\! \underbrace{\begin{bmatrix}
0 & \!\!\!\!H_{2,j,m}^{\mathrm{P}} \\
H_{2,j,m}^{\mathrm{P}} & \!\!\!\!0 \\
\end{bmatrix}}_{\triangleq \mathbf{H}_{2,j,m}^{\mathrm{P}}},
\end{align}
where $H_{1,j,m}^{\mathrm{P}} \triangleq \frac{\partial^{2}\bar{\mathsf{P}}_{j}(\mathbf{t}_{\mathrm{B},m})}{\partial[\mathbf{t}_{\mathrm{B},m}]_{1}^{2}}$, $H_{2,j,m}^{\mathrm{P}} \triangleq \frac{\partial^{2}\bar{\mathsf{P}}_{j}(\mathbf{t}_{\mathrm{B},m})}{\partial[\mathbf{t}_{\mathrm{B},m}]_{1}\partial[\mathbf{t}_{\mathrm{B},m}]_{2}}$, and $H_{3,j,m}^{\mathrm{P}} \triangleq \frac{\partial^{2}\bar{\mathsf{P}}_{j}(\mathbf{t}_{\mathrm{B},m})}{\partial[\mathbf{t}_{\mathrm{B},m}]_{2}^{2}}$. Then, the construction for $\mathbf{M}_{j,m}^{\mathrm{P}}$ can be transformed into deriving the lower bounds of $\mathbf{H}_{1,j,m}^{\mathrm{P}}$ and $\mathbf{H}_{2,j,m}^{\mathrm{P}}$, respectively.

Considering $\mathbf{H}_{1,j,m}^{\mathrm{P}}$ first, its lower bound can be achieved by exploiting the lower bounds of $H_{1,j,m}^{\mathrm{P}}$ and $H_{3,j,m}^{\mathrm{P}}$, respectively. Utilizing the fact that $\mathsf{cos}(\cdot) \in [-1, 1]$, we obtain
\begin{align}
H_{1,j,m}^{\mathrm{P}} &\geq -2{\sum}_{l=1}^{L_{\mathrm{G}}^{\mathrm{t}}}|[\mathbf{b}_{1,j,m}]_{l}|\frac{4\pi^{2}}{\lambda_{\mathrm{c}}^{2}}[\bm{\rho}_{\mathrm{G},l}^{\mathrm{t}}]_{1}^{2}, \\
H_{3,j,m}^{\mathrm{P}} &\geq -2{\sum}_{l=1}^{L_{\mathrm{G}}^{\mathrm{t}}}|[\mathbf{b}_{1,j,m}]_{l}|\frac{4\pi^{2}}{\lambda_{\mathrm{c}}^{2}}[\bm{\rho}_{\mathrm{G},l}^{\mathrm{t}}]_{2}^{2}.
\end{align}
Hence, via surrogating the diagonal elements of $\mathbf{H}_{1,j,m}^{\mathrm{P}}$ by (75) and (76), respectively, a lower bound of $\mathbf{H}_{1,j,m}^{\mathrm{P}}$ can be acquired.

Next, we explore a lower bound for $\mathbf{H}_{2,j,m}^{\mathrm{P}}$. Note that the following inequality
\begin{align}
\mathbf{H}_{2,j,m}^{\mathrm{P}} \geq -|H_{2,j,m}^{\mathrm{P}}|\mathbf{I}_{2}
\end{align}
naturally stands. Therefore, we need to examine the upper bound of $|H_{2,j,m}^{\mathrm{P}}|$. Resulting from $\mathsf{cos}(\cdot) \in [-1, 1]$, $|H_{2,j,m}^{\mathrm{P}}|$ can be upper-bounded by
\begin{align}
|H_{2,j,m}^{\mathrm{P}}| \leq 2{\sum}_{l=1}^{L_{\mathrm{G}}^{\mathrm{t}}}|[\mathbf{b}_{1,j,m}]_{l}|\frac{4\pi^{2}}{\lambda_{\mathrm{c}}^{2}}\big|[\bm{\rho}_{\mathrm{G},l}^{\mathrm{t}}]_{1}[\bm{\rho}_{\mathrm{G},l}^{\mathrm{t}}]_{2}\big|.
\end{align}
Consequently, by substituting (78) into (77), a lower bound of $\mathbf{H}_{2,j,m}^{\mathrm{P}}$ can be yielded, and the task of constructing $\mathbf{M}_{j,m}^{\mathrm{P}}$ is thus accomplished.


\vspace{-2mm}
\begin{thebibliography}{99}
\bibitem{ref1}Y. Zeng, B. Clerckx, and R. Zhang, “Communications and signals design for wireless power transmission,” \emph{IEEE Trans. Commun.}, vol. 65, no. 5, pp. 2264–2290, May 2017.
\bibitem{ref2}E. G. Larsson, O. Edfors, F. Tufvesson, and T. L. Marzetta, “Massive MIMO for next generation wireless systems,” \emph{IEEE Commun. Mag.}, vol. 52, no. 2, pp. 186–195, Feb. 2014.
\bibitem{ref3}Q. Wu, G. Y. Li, W. Chen, D. W. K. Ng, and R. Schober, “An overview of sustainable green 5G networks,” \emph{IEEE Wireless Commun.}, vol. 24, no. 4, pp. 72–80, Aug. 2017.
\bibitem{ref4}Q. Wu et al., “Intelligent surfaces empowered wireless network: Recent advances and the road to 6G,” \emph{Proc. IEEE}, vol. 112, no. 7, pp. 724–763, Jul. 2024.
\bibitem{ref5}Q. Wu and R. Zhang, “Intelligent reflecting surface enhanced wireless network via joint active and passive beamforming,” \emph{IEEE Trans. Wireless Commun.}, vol. 18, no. 11, pp. 5394–5409, Nov. 2019.
\bibitem{ref6}C. Pan et al., “Multicell MIMO communications relying on intelligent reflecting surfaces,” \emph{IEEE Trans. Wireless Commun.}, vol. 19, no. 8, pp. 5218–5233, Aug. 2020.
\bibitem{ref7}C. Huang, A. Zappone, G. C. Alexandropoulos, M. Debbah, and C. Yuen, “Reconfigurable intelligent surfaces for energy efficiency in wireless communication,” \emph{IEEE Trans. Wireless Commun.}, vol. 18, no. 8, pp. 4157–4170, Aug. 2019.
\bibitem{ref8}Q. Wu and R. Zhang, “Weighted sum power maximization for intelligent reflecting surface aided SWIPT,” \emph{IEEE Wireless Commun. Lett.}, vol. 9, no. 5, pp. 586–590, May 2020.
\bibitem{ref9}Q. Wu and R. Zhang, “Joint active and passive beamforming optimization for intelligent reflecting surface assisted SWIPT under QoS constraints,” \emph{IEEE J. Sel. Areas Commun.}, vol. 38, no. 8, pp. 1735–1748, Aug. 2020.
\bibitem{ref10}Y. Gao, Q. Wu, W. Chen, C. Wu, D. W. K. Ng, and N. A.-Dhahir, “Exploiting intelligent reflecting surfaces for interference channels with SWIPT,” \emph{IEEE Trans. Wireless Commun.}, vol. 23, no. 5, pp. 4442–4458, May 2024.
\bibitem{ref11}S. Zargari, A. Khalili, Q. Wu, M. R. Mili, and D. W. K. Ng, “Max-min fair energy-efficient beamforming design for intelligent reflecting surface-aided SWIPT systems with non-linear energy harvesting model,” \emph{IEEE Trans. Veh. Technol.}, vol. 70, no. 6, pp. 5848–5864, Jun. 2021.
\bibitem{ref12}Z. Zhu et al., “Reinforcement learning based resource allocation in IRS assisted SWIPT systems,” \emph{IEEE Trans. Veh. Technol.}, vol. 74, no. 4, pp. 6790–6794, Apr. 2025.
\bibitem{ref13}L. Zhu et al., “A tutorial on movable antennas for wireless networks,” \emph{IEEE Commun. Surveys Tuts.}, early access, 2025.
\bibitem{ref14}K.-K. Wong, A. Shojaeifard, K.-F. Tong, and Y. Zhang, “Fluid antenna systems,” \emph{IEEE Trans. Wireless Commun.}, vol. 20, no. 3, pp. 1950–1962, Mar. 2021.
\bibitem{ref15}L. Zhu, W. Ma, and R. Zhang, “Modeling and performance analysis for movable antenna enabled wireless communications,” \emph{IEEE Trans. Wireless Commun.}, vol. 23, no. 6, pp. 6234–6250, Jun. 2024.
\bibitem{ref16}W. Ma, L. Zhu, and R. Zhang, “MIMO capacity characterization for movable antenna systems,” \emph{IEEE Trans. Wireless Commun.}, vol. 23, no. 4, pp. 3392–3407, Apr. 2024.
\bibitem{ref17}Y. Zhu, Q. Wu, Y. Liu, Q. Shi, and W. Chen, “Suppressing beam squint effect for near-field wideband communication through movable antennas,” \emph{IEEE Trans. Veh. Technol.}, early access, 2025.
\bibitem{ref18}P. Chen, Y. Yang, B. Lyu, Z. Yang, and A. Jamalipour, “Movable-antenna-enhanced wireless-powered mobile-edge computing systems,” \emph{IEEE Internet Things J.}, vol. 11, no. 21, pp. 35505–35518, Nov. 2024.
\bibitem{ref19}Y. Gao, Q. Wu, and W. Chen, “Movable antennas enabled wireless-powered NOMA: Continuous and discrete positioning designs,” \emph{arXiv preprint arXiv:2409.20485}, 2024.
\bibitem{ref20}Z. Huang, N. Li, and P. Wu, “Weighted sum power maximization for movable antenna assisted SWIPT networks,” \emph{IEEE Commun. Lett.}, early access, 2025.
\bibitem{ref21}X. Dong, W. Lyu, R. Yang, Y. Xiu, W. Mei, and Z. Zhang, “Movable antenna enhanced secure simultaneous wireless information and power transfer,” \emph{IEEE Commun. Lett.}, vol. 29, no. 10, pp. 2356–2360, Oct. 2025.
\bibitem{ref22}Q. Wu et al., “Integrating movable antennas and intelligent reflecting surfaces (MA-IRS): Fundamentals, practical solutions, and opportunities,” \emph{arXiv preprint arXiv:2506.14636}, 2025.
\bibitem{ref23}X. Wei, W. Mei, Q. Wu, Q. Jia, B. Ning, and Z. Chen, “Movable antennas meet intelligent reflecting surface: Friends or foes?,” \emph{IEEE Trans. Commun.}, early access, 2025.
\bibitem{ref24}Y. Sun, H. Xu, B. Ning, Z. Cheng, C. Ouyang, and H. Yang, “Sum-rate optimization for RIS-aided multiuser communications with movable antennas,” \emph{IEEE Wireless Commun. Lett.}, vol. 14, no. 2, pp. 450–454, Feb. 2025.
\bibitem{ref25}Y. Gao, Q. Wu, W. Mei, G. Chen, W. Chen, and Z. Zheng, “Integrating movable antennas and intelligent reflecting surfaces for coverage enhancement,” \emph{arXiv preprint arXiv:2506.21375}, 2025.
\bibitem{ref26}Y. Ma, K. Liu, Y. Liu, L. Zhu, and Z. Xiao, “Movable-antenna aided secure transmission for RIS-ISAC systems,” \emph{IEEE Trans. Wireless Commun.}, early access, 2025.
\bibitem{ref27}Y. Geng, T. H. Cheng, K. Zhong, K. C. Teh, and Q. Wu, “Joint beamforming and antenna position optimization for IRS-aided multi-user movable antenna systems,” \emph{IEEE Trans. Wireless Commun.}, early access, 2025.
\bibitem{ref28}Y. Zhu, Q. Wu, W. Chen, Y. Liu, and R. Liu, “A flexible design for beam squint effect suppression in IRS-aided THz communications,” \emph{arXiv preprint arXiv:2508.21295}, 2025.
\bibitem{ref29}Q. Shi, M. Razaviyayn, Z.-Q. Luo, and C. He, “An iteratively weighted MMSE approach to distributed sum-utility maximization for a MIMO interfering broadcast channel,” \emph{IEEE Trans. Signal Process.}, vol. 59, no. 9, pp. 4331–4340, Sep. 2011.
\bibitem{ref30}D. Bertsekas, \emph{Nonlinear Programming}, 2nd ed. Belmont, MA, USA: Athena Scientific, 1999.
\bibitem{ref31}G. Chen, Q. Wu, C. He, W. Chen, J. Tang, and S. Jin, “Active IRS aided multiple access for energy-constrained IoT systems,” \emph{IEEE Trans. Wireless Commun.}, vol. 22, no. 3, pp. 1677–1694, Mar. 2023.
\bibitem{ref32}Q. Shi and M. Hong, “Penalty dual decomposition method for nonsmooth nonconvex optimization—Part I: Algorithms and convergence analysis,” \emph{IEEE Trans. Signal Process.}, vol. 68, pp. 4108–4122, Jun. 2020.
\bibitem{ref33}B. Zheng, C. You, W. Mei, and R. Zhang, “A survey on channel estimation and practical passive beamforming design for intelligent reflecting surface aided wireless communications,” \emph{IEEE Commun. Surveys Tuts.}, vol. 24, no. 2, pp. 1035–1071, 2nd Quart., 2022.
\bibitem{ref34}Y. Zhu, Y. Liu, Q. Wu, C. You, and Q. Shi, “Channel estimation by transmitting pilots from reconfigurable intelligent surface,” \emph{IEEE Trans. Wireless Commun.}, vol. 23, no. 4, pp. 3328–3343, Apr. 2024.
\bibitem{ref35}W. Ma, L. Zhu, and R. Zhang, “Compressed sensing based channel estimation for movable antenna communications,” \emph{IEEE Commun. Lett.}, vol. 27, no. 10, pp. 2747–2751, Oct. 2023.
\bibitem{ref36}Z. Xiao, S. Cao, L. Zhu, Y. Liu, B. Ning, and X.-G. Xia, “Channel estimation for movable antenna communication systems: A framework based on compressed sensing,” \emph{IEEE Trans. Wireless Commun.}, vol. 23, no. 9, pp. 11814–11830, Sep. 2024.
\bibitem{ref37}M. Grant and S. Boyd, “CVX: MATLAB software for disciplined convex programming,” 2016. [Online]. Available: http://cvxr.com/cvx
\bibitem{ref38}M. Razaviyayn, M. Hong, and Z.-Q. Luo, “A unified convergence analysis of block successive minimization methods for nonsmooth optimization,” \emph{SIAM J. Optim.}, vol. 23, no. 2, pp. 1126–1153, Jan. 2013.
\end{thebibliography}
\end{document}